\documentclass{article} 
\usepackage{iclr2025_conference}
\usepackage{times}
\iclrfinalcopy

\usepackage{amsmath,amsfonts,bm}

\def\eqref#1{equation~\ref{#1}}

\def\1{\bm{1}}

\DeclareMathAlphabet{\mathsfit}{\encodingdefault}{\sfdefault}{m}{sl}
\SetMathAlphabet{\mathsfit}{bold}{\encodingdefault}{\sfdefault}{bx}{n}

\usepackage{listings}
\usepackage{xcolor}
\usepackage{multirow}
\definecolor{codegreen}{rgb}{0,0.6,0}
\definecolor{codegray}{rgb}{0.5,0.5,0.5}
\definecolor{codepurple}{rgb}{0.58,0,0.82}
\definecolor{backcolour}{rgb}{0.95,0.95,0.92}

\lstdefinelanguage{toml}{
    comment=[l]{\#},
    keywords={true,false},
    morestring=[b]",
    morestring=[b]',
    sensitive=true,
}
\usepackage{tcolorbox}
\usepackage[compact]{titlesec}
\usepackage{longtable}
\usepackage{xspace}
\usepackage{enumitem}
\usepackage{hyperref}
\usepackage{url}
\usepackage{graphicx}
\usepackage{booktabs} 
\usepackage{stmaryrd}
\usepackage{subcaption}
\usepackage{tcolorbox}
\usepackage{xcolor}
\usepackage{graphicx}
\usepackage{comment}
\usepackage{amssymb}
\usepackage{amsmath}
\usepackage{bbm}
\newcommand{\tool}{QLCoder\xspace}

\newcommand{\Fix}[1]{\textcolor{red}{[[#1]]}}

\newcommand{\mypara}[1]{\vspace{.03in}\noindent \textbf{#1.}}

\newcommand{\ccomment}[1]{}
\newcommand{\CodeIn}[1]{{\small\ifmmode{\mathtt{#1}}\else$\mathtt{#1}$\fi}}
\newcommand{\changed}[1]{\textcolor{black}{#1}}
\newcommand{\datachange}[1]{\textcolor{black}{#1}}
\title{\tool: A Query Synthesizer for Static Analysis of Security Vulnerabilities}
\author{Claire Wang \\ University of Pennsylvania \\
\texttt{cdwang@seas.upenn.edu}
\And 
Ziyang Li \\ John Hopkins University \\ 
\texttt{ziyang@cs.jhu.edu}
\And 
Saikat Dutta \\ Cornell University \\ 
\texttt{saikatd@cornell.edu}
\And 
Mayur Naik \\ University of Pennsylvania \\ 
\texttt{mhnaik@upenn.edu}
}
\begin{document}
\maketitle
\begin{abstract}

Static analysis tools provide a powerful means to detect security vulnerabilities by specifying \emph{queries} that encode vulnerable code patterns. However, writing such queries is challenging and requires diverse expertise in security and program analysis. To address this challenge, we present \emph{\tool} -- an agentic framework that automatically synthesizes queries in CodeQL, a powerful static analysis engine, directly from a given CVE metadata. \tool embeds an LLM in a synthesis loop with execution feedback, while constraining its reasoning using a custom MCP interface that allows structured interaction with a Language Server Protocol (for syntax guidance) and a RAG database (for semantic retrieval of queries and documentation). This approach allows \tool to generate syntactically and semantically valid security queries.
%
We evaluate \tool on 176 existing CVEs across 111 Java projects. Building upon the Claude Code agent framework, \tool synthesizes correct queries that detect the CVE in the vulnerable but not in the patched versions for 53.4\% of CVEs. In comparison, using only Claude Code synthesizes 10\% correct queries. 
Our generated queries achieve an F1 score of  0.7. In comparison, the general query suites in IRIS (a recent LLM-assisted static analyzer) and CodeQL only achieve  F1 scores of 0.048 and 0.073, highlighting the benefit of \tool's specialized synthesized queries. QLCoder is available at \url{https://github.com/neuralprogram/qlcoder}.

\end{abstract}

\section{Introduction}
\vspace{-.1in}
Security vulnerabilities continue to grow at an unprecedented rate, with over 40,000 Common Vulnerabilities
and Exposures (CVEs) reported in 2024, 28,961 CVEs reported in 2023, and 25,059 CVEs reported in 2022 \citep{cve2025a}.
Static analysis, a technique to analyze programs without executing them, is a common way of detecting
vulnerabilities.
Static analysis tools such as CodeQL \citep{codeql2025}, Semgrep \citep{semgrep2023}, and Infer \citep{infer} are widely used in industry.
They provide domain-specific languages that allow specifying vulnerability patterns as queries. Such queries can be executed over structured representations of code, such as abstract syntax trees, to detect potential security vulnerabilities.

\begin{figure}[t] 
    \centering
    \includegraphics[width=\textwidth]{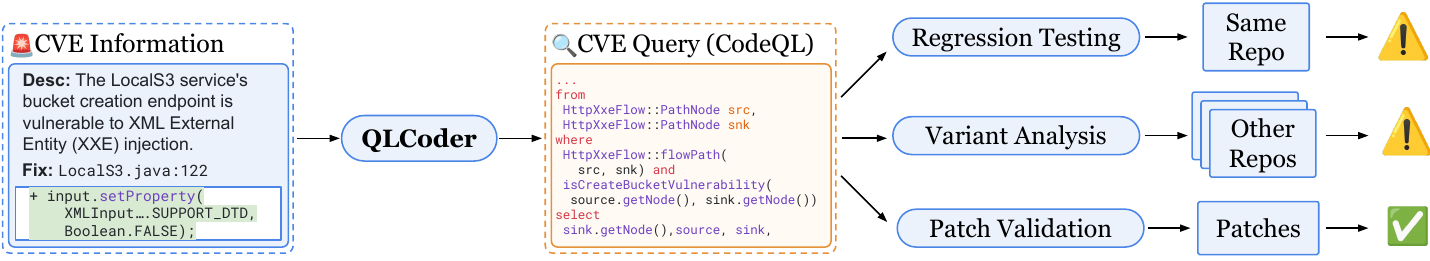}
    \vspace{-15px}
    \caption{
    A CodeQL query capturing a vulnerability pattern is synthesized by \tool from an existing CVE and subsequently reused for regression testing, variant analysis, or patch validation.
    }
    \label{fig:cve-query}
    \vspace{-0.2in}
\end{figure}

Despite their widespread use, existing query suites of static analysis tools are severely limited in coverage of vulnerabilities and precision.
Extending them is difficult even for experts, as it requires knowledge of unfamiliar query languages,
program analysis concepts, and security expertise.
Incorrect queries can produce false alarms or miss bugs, limiting the effectiveness of static analysis.
\ccomment{Conversely, }
Correct queries can enable reliable detection of real vulnerabilities, supporting diverse use-cases such as regression testing, variant analysis, and patch validation, among others (Figure~\ref{fig:cve-query}).

Meanwhile, CVE databases \citep{mitre,nvd,github-advisory} provide rich information about security vulnerabilities, including natural
language descriptions of vulnerability patterns and records of buggy and patched versions of the affected software repositories.
This resource remains largely untapped in the automated construction of static analysis queries.
Recent advances in LLMs, particularly in code understanding and generation, open up the possibility of leveraging this
information to automatically synthesize queries from  CVE descriptions, thereby bridging the gap between
vulnerability reports and practical detection tools.

Synthesizing such queries poses significant challenges.
The syntax of static analysis query languages is low-resource, richly expressive, and evolves continually.
A typical query, such as the one in Figure~\ref{fig:main-example}(b) specifying a global dataflow pattern leaves ample room for errors in describing predicates for sources, sinks, sanitizers, and taint propagation steps.
Even if the generated syntax is correct, success is measured by whether the query can identify at least one execution path traversing the bug location in the vulnerable version while producing no matches in the patched version. 
Achieving this requires understanding the CVE context at the level of abstract syntax trees, such as code differences that introduce a sanitizer to prevent a flow from a source to a sink. 
Complicating matters further, reasoning about the code changes alone is often insufficient: sources, sinks, and taint propagation steps may reside in parts of the codebase far from the modified functions or files, and the vulnerability itself may involve non-trivial dataflow chains across these components. 
Thus, a correct query must not only integrate information from multiple locations across the program but also capture the intricate propagation patterns to accurately characterize the vulnerability.

In this paper, we present \textbf{\tool{}}--an agentic framework that synthesizes queries in CodeQL, a powerful static analysis engine, directly from a given CVE metadata.
We select CodeQL because it has the richest query language, which allows capturing complex inter-procedural vulnerability patterns. 
\tool{} addresses the above challenges by embedding an LLM in a structured synthesis loop that incorporates execution feedback to verify query correctness and allows interactive reasoning using a custom MCP (Model Context Protocol) interface.
The MCP interface constrains the model's reasoning using a Language Server Protocol (for syntax guidance) and a vector database of CodeQL queries and documentation (for semantic guidance). 
By combining these capabilities, \tool{} avoids common pitfalls of naive LLM-based approaches, such as producing ill-formed queries, hallucinating deprecated constructs, or missing subtle vulnerability patterns, and instead produces queries that are both syntactically correct and semantically precise.

We evaluate \tool{} on CWE-Bench-Java~\citep{li2025iris}, which comprises 176 CVEs across 111 Java projects.
These CVEs span 42 different Common Weakness Enumeration (CWE) categories and the projects range in size from 0.01 to 1.5 MLOC.
To account for model training cut-offs, we include 65 CVEs reported during 2025 and target a recent CodeQL version 2.22.2 (July 2025).
Using the Claude Code agent framework, \tool{} achieves query compilation and success rates of \datachange{99.4}\% and 53.4\%, compared to \datachange{60}\% and 0\% for our best agentic baseline, Gemini CLI.
Further, our generated queries have an F1 score of 0.7 for detecting true positive vulnerabilities, compared to 0.048 for
IRIS~\citep{li2025iris}, a recent LLM-assisted static analyzer, and  0.073 for CodeQL.

We summarize our main contributions:
\begin{itemize}[leftmargin=*,noitemsep,topsep=0pt]
\item \textbf{Agentic Framework for CVE-to-Query Synthesis.} We present \tool{}, an agentic framework that translates CVE descriptions into executable CodeQL queries, bridging the gap between vulnerability reports and static analysis. \tool{} introduces a novel integration of execution-guided synthesis, semantic retrieval, and structured reasoning for vulnerability query generation.
\item \textbf{Evaluation on Real-World Repositories and CVEs.}
We evaluate \tool on 176 CVEs in Java projects, covering 42 vulnerability types (CWEs) from CWE-Bench-Java. Each project involves complex inter-procedural vulnerabilities spanning multiple files. We show how \tool can successfully identify sources, sinks, sanitizers, and taint propagation steps, and refine queries to ensure they raise alarms on vulnerable versions while remaining silent on patched versions.
\item \textbf{Comparison with Baselines.} We compare \tool against state-of-the-art agent frameworks and show that \tool achieves substantially higher compilation, success, and F1 scores. We also compare \tool's synthesized queries with state-of-the-art static analysis frameworks and show that our queries are more precise and have higher recall.
\end{itemize}

\section{Illustrative Example}
\label{sec:motivating}

\begin{figure}[t]
    \begin{subfigure}{\linewidth}
        \includegraphics[width=\linewidth]{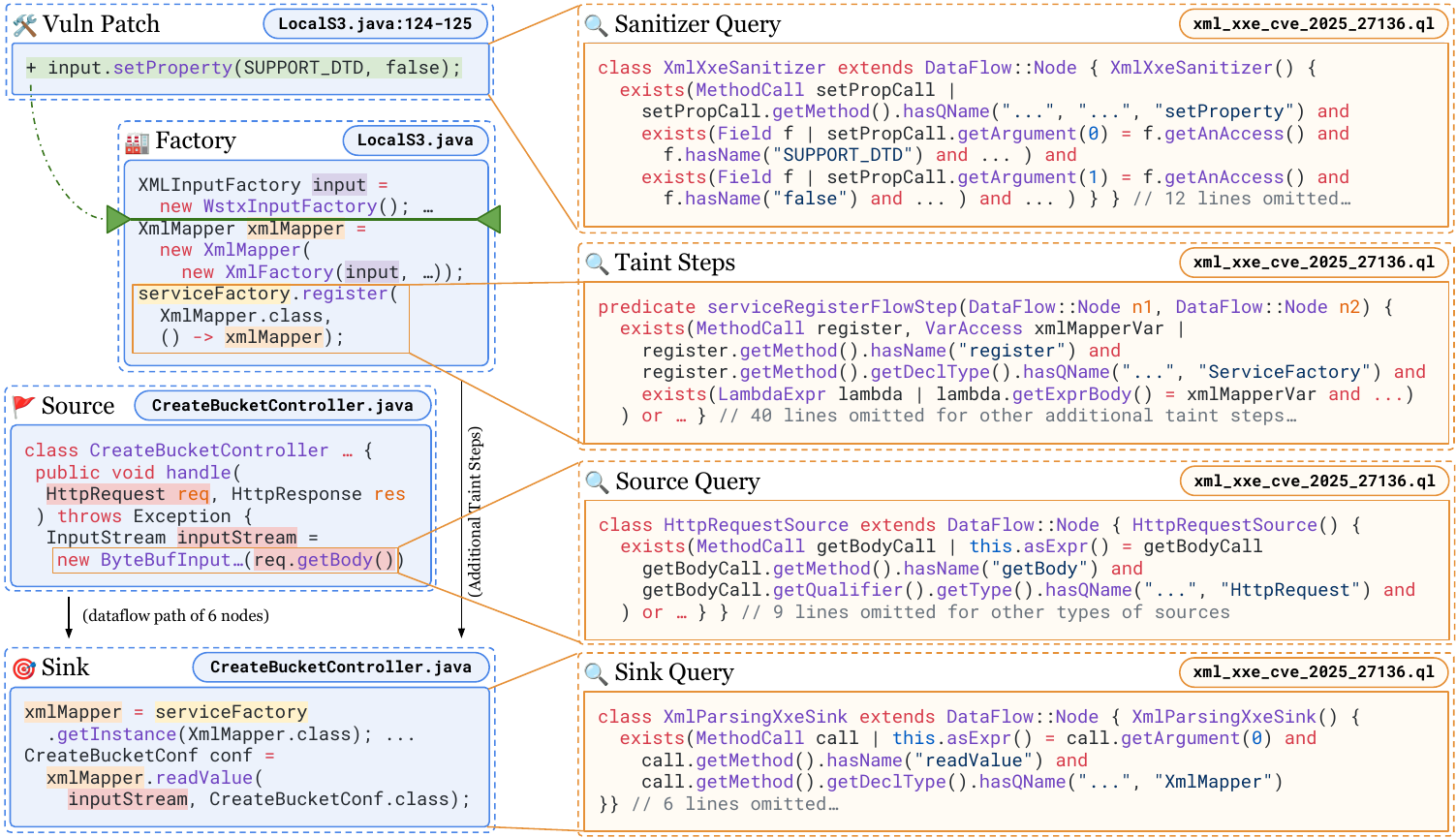}
    \end{subfigure}

    \begin{subfigure}{0.34\linewidth}
        \caption{
            The vulnerable dataflow snippets and the patch, which adds configuration to \texttt{XMLInputFactory}.
        }
        \label{fig:motivating-vul-snippets}
    \end{subfigure}
    \hfill
    \begin{subfigure}{0.60\linewidth}
        \caption{
            Snippets of the synthesized vulnerability query by \tool capturing the patterns of dataflow source, sink, taint steps, and the sanitizer indicated by the vulnerability patch.
        }
        \label{fig:motivating-query-components}
    \end{subfigure}
    \begin{subfigure}{\linewidth}
    \vspace{0.15in}
        \includegraphics[width=\linewidth]{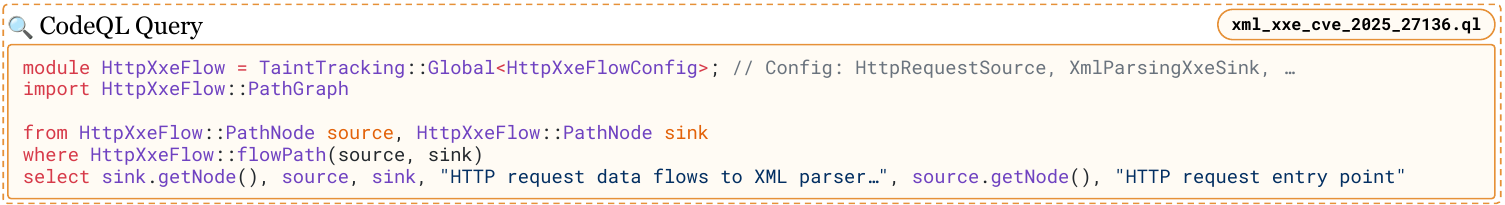}
        \caption{
            The synthesized CodeQL path query that ties everything together.
        }
        \label{fig:motivating-query}
    \end{subfigure}
    
    \caption{
        Illustration of vulnerability CVE-2025-27136 in repository \texttt{Robothy/local-s3} which exhibits an XML External Entity Injection weakness (CWE-611).
        When the \texttt{XmlMapper} is not configured to disable \textit{Document Type Definition} (DTD), the function \texttt{readValue} may declare additional entities, allowing hackers to inject malicious behavior.
    }
    \label{fig:main-example}
\end{figure}

We illustrate the challenges of vulnerability query synthesis using CVE-2025-27136, an XML External Entity Injection (XXE) bug found in the repository \texttt{Robothy/local-s3}.
Figure~\ref{fig:main-example} depicts the vulnerability snippets, the patch, and the synthesized CodeQL query generated by \tool.

\mypara{Vulnerability context}
The vulnerability arises when the \texttt{XmlMapper} object is used to parse user-provided XML data (Figure~\ref{fig:motivating-vul-snippets}).
In the vulnerable code, \texttt{XmlMapper.readValue} is called on the HTTP request body without disabling support for \textit{Document Type Definitions} (DTDs).
As a result, an attacker can inject malicious external entity declarations into the input stream, enabling server-side request forgery (SSRF) attacks, allowing for access to resources that should not be accessible from external networks, effectively leaking sensitive information.
The patch mitigates the issue by configuring the underlying \texttt{XMLInputFactory} with the property \texttt{SUPPORT\_DTD=false}.

\mypara{Synthesizing the query}
The CodeQL query that can effectively capture the vulnerability pattern needs to incorporate
1) sources such as \texttt{HttpRequest.getBody} calls where untrusted malicious information enters the program,
2) sinks such as invocations of \texttt{XmlMapper.readValue}, where the XXE vulnerability is manifested,
3) additional taint steps related to how the \texttt{XmlMapper} is constructed and configured, involving non-trivial interprocedural flows spanning multiple files,
and
4) sanitizers such as calls to \texttt{setProperty(SUPPORT\_DTD, false)}, so that we know that no alarm should be reported after the vulnerability has been fixed.

In general, the synthesized query must connect all these components to be able to detect the bug in the vulnerable program, while not reporting the same alarm after the vulnerability has been fixed.
Figure~\ref{fig:motivating-query-components} shows all the components of the CodeQL query (simplified), capturing their individual syntactic patterns.
Lastly, Figure~\ref{fig:motivating-query} connects all these components into a coherent path query by using CodeQL's \texttt{TaintTracking::Global<.>::PathGraph} and the SQL-like \texttt{from-where-select} query, which returns the exact path from source to sink.

\mypara{Challenges and solutions}
Vulnerability query synthesis must overcome several tightly-coupled challenges.
We hereby state the challenges and explain how \tool addresses them.
\begin{itemize}[leftmargin=*,itemsep=1px]
    \item \textit{Rich expressiveness and fragility of syntactic patterns.}
    CodeQL is powerful but syntactically intricate: small mistakes in predicate names, qualifiers, or AST navigation often produce syntactically valid yet semantically useless queries.
    \tool mitigates this fragility through its Language Server Protocol (LSP) interface for syntax guidance and RAG database for semantic retrieval of existing CodeQL queries and documentation.
    These structured interactions guide predicate selection and AST navigation during synthesis, reducing off-by-name and version-mismatch errors.


    \item \textit{Inter-procedural taint propagation across a large codebase.}
Sources, sinks, and sanitizers typically live in different modules or files and are connected by nontrivial inter-procedural flows (lambdas, factory patterns, etc.). While CodeQL provides robust inter-procedural analysis for many common patterns, gaps in dataflow still require bridging via additional taint propagation steps. Through its custom MCP interface, \tool performs structured reasoning to discover candidate program points, synthesize custom taint-step predicates (e.g., service registration), and compose them into a CodeQL path query that tracks data across file and component boundaries.


    \item \textit{Semantic precision: alarm on the vulnerable version, silence on the patched version.}
    A useful vulnerability query must not only parse correctly but also be discriminative.
    \tool enforces this semantic requirement directly during synthesis.
    Via an iterative refinement loop, the successful criteria states that in the fixed program, there should be no alarm being raised about the vulnerability.
    This incentivizes the agent to synthesize sanitizer predicates (e.g., the \texttt{setProperty} call) and use them to constrain the path query so that sanitizer presence suppresses the alarm.
    The resulting query thus captures the exact behavior difference, producing alarms on the vulnerable snapshot and not on the patched snapshot.
\end{itemize}

Together, these capabilities let \tool synthesize a semantically precise CodeQL query that can be reused for regression testing, variant analysis, or patch validation.
We now elaborate on the detailed design and implementation of \tool.

\section{\tool}
\vspace{-.1in}
\label{sec:methodology}

\begin{figure}[t]
    \centering
    \includegraphics[width=\linewidth]{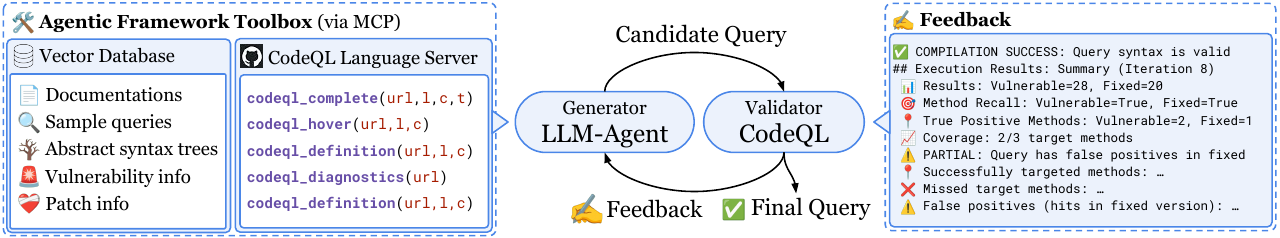}
    \caption{
        Overall pipeline of \tool's iterative synthesis loop between an agentic query generator and a CodeQL-based validator. 
        The generator uses a vector database and our CodeQL Language Server as tools while
        the validator produces compilation, execution, and coverage feedback.
    }
    \label{fig:iterative-refinement}
    \vspace{-0.15in}
\end{figure}

At a high level, \tool operates inside a repository-aware iterative refinement loop (Figure~\ref{fig:iterative-refinement}). 
In each iteration, the agent proposes a candidate CodeQL query, a CodeQL-based validator executes and scores it on both the vulnerable and patched versions of the repository, and the agent uses the validation feedback to propose targeted repairs.
The loop terminates successfully when the validator accepts a query, or fails after a fixed iteration budget. 
In this section, we elaborate the major design components that make the loop effective.

\subsection{Problem Statement}
\label{sec:problem-statement}
\vspace{-.1in}
The task of vulnerability detection is generally framed as a taint analysis task, where the goal of a \emph{query} is to find dataflow paths from a \emph{source} (e.g., an API endpoint accepting user input) to a \emph{sink} (e.g., a database write) that lack proper \emph{sanitization} (e.g., filtering malicious data).\ccomment{ -- also known as \emph{tainted paths}.}

We formalize the \textit{Vulnerability Query Synthesis} problem as follows. 
Assume as input a vulnerable project version \(P_{\mathrm{vuln}}\), 
its fixed version \(P_{\mathrm{fixed}}\), and 
a textual CVE description (commonly available in open vulnerability reports). 
Let us assume we have inter-procedural dataflow program graphs for each code version: 
\(G_{\mathrm{vuln}}=(V_{\mathrm{vuln}},E_{\mathrm{vuln}})\) and 
\(G_{\mathrm{fixed}}=(V_{\mathrm{fixed}},E_{\mathrm{fixed}})\).
Let \(\Delta P\) denote the source-level patch between \(P_{\mathrm{vuln}}\) and \(P_{\mathrm{fixed}}\).  
We represent the patch in the dataflow-graph domain as a patch subgraph
\(\Delta G=(\Delta V,\Delta E)\), where \(\Delta V\) is the set of graph nodes that correspond to the modified program snippets.

A \emph{vulnerability path query} \(Q\) evaluated on a graph \(G\) returns a set of dataflow paths, denoted as \(\Pi = \llbracket Q\rrbracket(G)\). 
We write each path $\pi \in \Pi$ as \(\pi=\langle v_1,\dots,v_k\rangle\), where each $v_i \in V$ is a node in the dataflow graph $G$. 
Consecutive nodes $(v_i, v_{i+1})$ should be either connected by an existing edge in $E$, or an \textit{additional taint step} specified in the query $Q$, to compensate for missing edges via dataflow graph construction.
Specifically, we call \(v_1\) the \emph{source} of path $\pi$ and \(v_k\) the \emph{sink} of $\pi$. 

\mypara{Synthesis task} We aim to synthesize a query \(Q\) from the vulnerability report satisfying the following requirements:
\begin{enumerate}[leftmargin=*,noitemsep,topsep=0pt]
  \item \textbf{Well-formedness.} \(Q\) is syntactically valid (based on the latest CodeQL syntax) and can be executed on the target CodeQL infrastructure (e.g., dataflow graphs) without runtime errors.
  \item \textbf{Vulnerability detection.} \(Q\) generates at least one path \(\pi\) in the vulnerable version that traverses the patched region:
  \[
    \exists \pi\in\llbracket Q\rrbracket(G_{\mathrm{vuln}})\quad\text{such that}\quad \pi\cap\Delta V \neq \emptyset.
  \]
  \item \textbf{Fix discrimination.} \(Q\) does not report the vulnerability in the fixed version. Concretely, no path reported on the fixed version should traverse the patched locations:
  \[
    \forall \pi\in\llbracket Q\rrbracket(G_{\mathrm{fixed}}),\quad\text{we have}\quad \pi \cap \Delta V=\emptyset.
  \]
\end{enumerate}
In other words, the synthesized query must be executable, must witness the vulnerability in the vulnerable version via a path that uses code touched by the fix, and must not attribute the same (patched) behavior in the fixed version.
When only the well-formedness condition is satisfied, we say that the query $Q$ is valid (denoted as \(\CodeIn{valid}(Q)\));
when all the conditions are satisfied, the query $Q$ is successful (denoted as \(\CodeIn{success}(Q; P_\mathrm{vuln}, P_\mathrm{fixed})\)). 
Note that these criteria may admit potentially false positive paths in both versions. It might be possible to consider additional constraints regarding precision, but it might further complicate synthesis. In practice, we find most queries synthesized by \tool already have high precision.

\subsection{Design of \tool}
\vspace{-.1in}
Concretely, \tool proceeds in an iterative refinement loop indexed by $i = [0,1,\dots ]$. 
Via prompting, the LLM agent-based synthesizer first proposes an initial candidate query \(Q_0\). 
For each iteration \(i\), the validator evaluates \(Q_i\) and produces a feedback report. 
We consider synthesis successful at iteration \(i\) iff \(\CodeIn{success}(Q_i;P_{\mathrm{vuln}},P_{\mathrm{fixed}})\) holds; 
in that case the loop terminates and \(Q_i\) is returned. 
Otherwise, the synthesizer analyzes the feedback and the previous candidate \(Q_i\), and produces the next query candidate \(Q_{i+1}\).
The loop stops successfully when \(\CodeIn{success}(\cdot)\) is achieved or fails once \(i\) reaches the pre-configured limit \(N\) (in our implementation \(N=10\)).
The remainder of the design focuses on two aspects: 
1) how the agentic synthesizer performs synthesis, and 
2) how the validator generates and communicates feedback. 
We elaborate on both below.

\mypara{Agentic synthesizer}
In each iteration $i$, the LLM-based agentic synthesizer runs an inner \emph{conversation loop} of up to  $M$ turns. 
In each turn, the agent either performs internal reasoning or issues a tool call by emitting a JSON-formatted action. 
When a tool call succeeds, the tool returns a JSON-formatted response that is appended to the conversation history. 
Conversation histories are kept local to the current refinement iteration (i.e.,  not carried over between iterations) to keep context compact and relevant. 
In practice, we set $M = 50$, i.e., the agent may interact with tools up to 50 times before generating a candidate query for validation. 

Two design choices are critical for the effectiveness of this loop: 
1) the \emph{initial prompt} that initializes and constrains the agent’s behavior, and 
2) the \emph{toolbox} of callable tools, each exposed by a custom Model Context Protocol (MCP) server. 
We refer to the combined problem of designing these items as \emph{Context Engineering} (discussed in Section~\ref{sec:context-engineering}).

\mypara{CodeQL-Based Validator}
The validator compiles and executes each candidate query against the vulnerable and fixed versions and returns a concise, structured feedback report that is used to drive refinement (Figure~\ref{fig:iterative-refinement}). 
The vulnerable and patched versions are automatically derived from patch commit hash, which is in the CVE metadata.
The report contains: 
(i) CodeQL compilation results, 
(ii) execution counts (matches on vulnerable and fixed graphs), 
(iii) recall and coverage statistics, 
(iv) concrete counterexample traces and hit locations, and 
(v) a prioritized set of next-step recommendations (e.g., add qualifiers, synthesize sanitizer checks, or expand taint steps) that are programmatically generated via a template.

\subsection{Context Engineering for Agentic Synthesizer}
\label{sec:context-engineering}
\begin{figure}[t]
    \includegraphics[width=\linewidth]{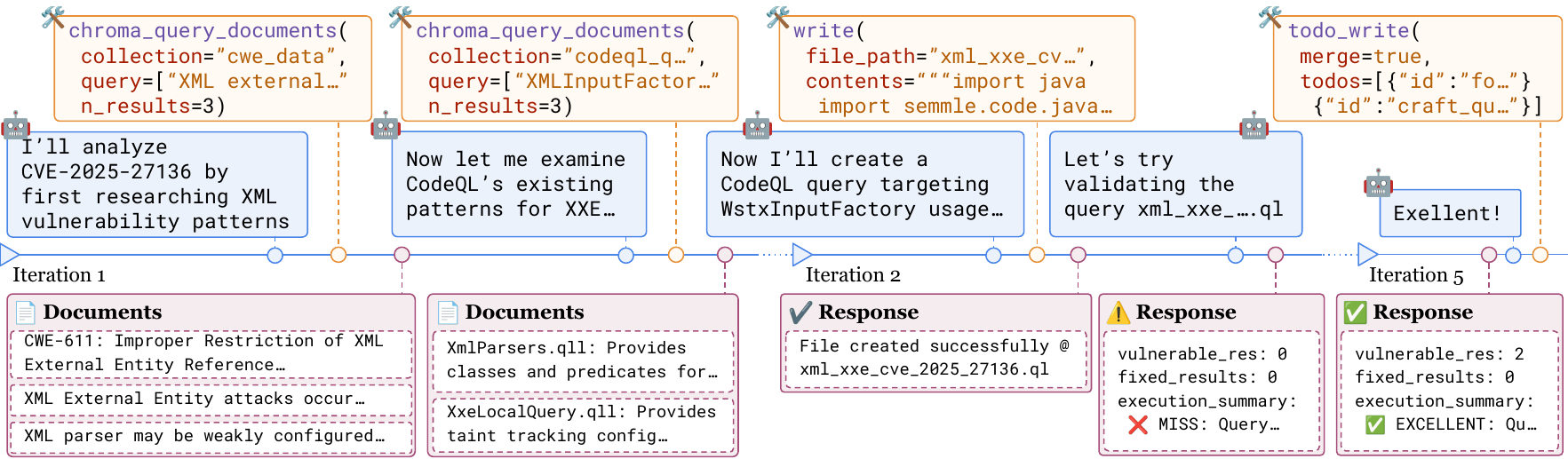}
    \caption{
        Illustration of example traces of conversation during the synthesis of the query in the motivating example (Figure~\ref{fig:main-example}).
        LLM-agent may think, invoke tools that are available in the toolbox, and receive responses from the MCP servers.
    }
    \label{fig:refinement-conversation}
    \vspace{-0.15in}
\end{figure}

\vspace{-.1in}
The primary goal of context engineering is to expose the LLM-based agent to the most \emph{precise} amount of information: 
enough for the agent to make progress, but not so much that the LLM is confused or the cost explodes.
As illustrated in Figure~\ref{fig:iterative-refinement}, \tool relies on two primary MCP servers to provide demand-driven, structured information to the agent: 
a retrieval-augmented vector database and a CodeQL Language Server interface.
We show example traces of conversation loop in Figure~\ref{fig:refinement-conversation} and describe the available tools below.

\mypara{Initial prompt}
Each refinement iteration begins with an \emph{initial prompt} that kicks-off the agentic conversation loop. 
The initial prompt in the first iteration contains a query skeleton for reference 
(See \S~\ref{sec:appendix-query-structure} for an example).
In subsequent iterations, the prompt contains a summary of the synthesis goal and constraints, the previous candidate query \(Q_{i-1}\), and the validator feedback report. 
Concretely, the initial prompt emphasizes: 
(i) the success predicate (see \(\CodeIn{Success}(\cdot)\)), 
(ii) concrete counterexamples from previous feedback, and 
(iii) an explicit list of callable tools and their purpose.

\mypara{Vector database}
We use a retrieval-augmented vector database (ChromaDB MCP server in our implementation) to store large reference corpora without polluting the LLM prompt. 
The database is pre-populated with 
(i) vulnerability analysis notes and diffs, 
(ii) Common Weaknesses Enumeration (CWE) definitions, 
(iii) same-version CodeQL API documentation, 
(iv) curated CodeQL sample queries, and 
(v) small abstract syntax tree (AST) snippets extracted from the target repository. \changed{We manually filtered the patch diffs files to be only related to the given CVE and automated removing non-Java files that couldn't be analyzed by CodeQL or were unrelated to the CVE pattern, such as binary files. Afterwards, ASTs of the related patch files were extracted.}
During a conversation loop, the agent issues compact retrieval queries (e.g., to fetch example CodeQL queries related to the CWE) 
and receives ranked documents or snippets on demand. 

In practice, we may populate our RAG database with tens of thousands of documents. Even with this large corpus, we observe that the LLM-agent reliably retrieves exactly the kinds of artifacts it needs: 
CodeQL sample queries that inspire overall query structure, 
small AST snippets that suggest the precise syntactic navigation, and vulnerability writeups or diff excerpts that help discriminate buggy from patched behavior. 
These demand-driven lookups let the agent gather high-quality information without loading the main prompt with large reference corpora.

\mypara{CodeQL language server}
We expose the CodeQL Language Server \citep{noauthor_execute_nodate} through a MCP server that the agent can call for precise syntax-aware guidance. Importantly, we developed our own CodeQL Language Server client and MCP server that ensures syntactic validity (especially for the given CodeQL version) during query generation. The LLM agent's MCP client makes the tool call which is received by the CodeQL MCP server.
The MCP server forwards tool calls, such as \texttt{complete(file, loc, char)}, \texttt{diagnostics(file)}, and \texttt{definition(file, loc, char)}, to the underlying CodeQL process and returns JSON-serializable responses. 
Tools such as \texttt{completion} help the agent fill query templates and discover correct API or AST names, 
while \texttt{diagnostics} reveal compile or linter errors (e.g., unknown predicate names) that guide mutation.
Appendix~\ref{sec:appendix-codeql-lsp} shows the full specification and example request and response schemas.
%

\vspace{-.3in}
\changed{\subsection{Design Decisions}}
\vspace{-.1in}
We discuss several alternative designs that we considered but found ineffective in practice.  
Allowing the agent unrestricted access to compile-and-run CodeQL via MCP led to severe performance degradation: 
compilation and full execution are expensive operations that the LLM soon overused, so we instead expose only lightweight diagnostics during the conversation and defer full compile-and-run to the end of each iteration. 
Permitting web search for vulnerability patterns or snippets similarly proved problematic. 
It is both costly and easy for the agent to rely on web lookups, which quickly pollutes the working context and degrades synthesis quality. \changed{See Appendix \ref{sec:appendix-design} for more details on using web search.}
Equipping the agent with an extensive set of heterogeneous tools led to confusion and poor tool-selection behavior; in contrast, a small, well-scoped toolbox yields more reliable actions.  
Finally, retaining full conversation histories across refinement iterations induced context rot and ballooning prompt sizes, so we keep histories local to each iteration.  
Overall, our current design is a pragmatic trade-off that balances cost, responsiveness, and synthesis effectiveness.

\section{Evaluation}
\vspace{-.1in}
We aim to answer the following research questions through our empirical evaluation:
\begin{itemize}[leftmargin=*,noitemsep,topsep=0pt,partopsep=0pt]
  \item \textbf{RQ 1}: For how many CVEs can \tool successfully generate queries? 
  \item \textbf{RQ 2}: How useful is each component of \tool? 
  \item \textbf{RQ 3:} How does the choice of base agent framework affect \tool's effectiveness?
  \item \textbf{RQ 4:} How scalable is \tool?
\end{itemize}
\vspace{-5px}
\subsection{Experimental Setup}
\vspace{-.1in}
We develop \tool on top of the Claude Code framework~\citep{claude-code2025} and use Claude Sonnet 4 for all our experiments. \datachange{For agent baselines, we used Codex with GPT-5 (minimal/medium reasoning) and Gemini CLI with Gemini 2.5 Flash/Pro.}
For each CVE and agent baseline, we use a maximum of 10 iterations ($N = 10$).
For static analysis baselines, we select IRIS~\cite{li2025iris}\ccomment{, which depends on an older version of CodeQL (2.15.*)} and CodeQL (version 2.22.2) query suites. 
Experiments were run on machines with the following specifications: for the gpt-oss agent baseline we used a machine with \changed{an Intel Xeon Gold 6338 2.00 GHz CPU, 10x NVIDIA H100 PCIe, and 1 TB RAM.} For all other experiments we used an Intel Xeon Gold 6248 2.50GHz CPU, four GeForce RTX 2080 Ti GPUs, and 750GB RAM. 

\mypara{Dataset} We used CWE-Bench-Java \citep{li2025iris} and its latest update, which added new CVEs from 2025. We were able to successfully build and use 111 (out of 120) Java CVEs evaluated in IRIS~\citep{li2025iris}, and 65 (out of 91) 2025 CVEs. 
Each sample in CWE-Bench-Java comes with the CVE metadata and fix commit information associated with the bug. 
\vspace{-5px}
\subsection{Evaluation Metrics}
\vspace{-.1in}
Besides $\CodeIn{valid}(Q)$ and $\CodeIn{success}(Q; P_\mathrm{vuln}, P_\mathrm{fixed})$  from Section \ref{sec:problem-statement}, we use the following terms and metrics when evaluating \tool and baselines on the problem of vulnerability query synthesis:
\begin{align*}
\footnotesize
\text{Rec}(Q) = \mathbbm{1}[\exists \pi \in \llbracket Q \rrbracket(G_\mathrm{vuln}), \pi \cap \Delta V \neq \emptyset],
&\quad
\text{Prec}(Q) = \frac{|\{\pi \in \llbracket Q \rrbracket (G_\mathrm{vuln}) ~|~ \pi \cap \Delta V \neq \emptyset \}|}{|\llbracket Q \rrbracket (G_\mathrm{vuln})|}, \\
\text{F1}(Q) = &2 \cdot \frac{\text{Prec}(Q) \cdot \text{Rec}(Q)}{\text{Prec}(Q) + \text{Rec}(Q)}.
\end{align*}

\vspace{-.1in}
\subsection{RQ1: \tool Effectiveness}
\vspace{-.1in}
\begin{table}[t]
\scriptsize 
    \caption{\tool Query Success by CWE Type.}
    \vspace{-10pt}
    \label{tab:cwe_success_results}
    \centering
    \begin{tabular}{@{}l@{\hspace{0.5em}}rrrr@{}}
    \toprule
    \textbf{CWE Type} & \textbf{Total CVEs} & \textbf{\# Success} & \textbf{Success (\%)} & \textbf{Avg Precision} \\
    \midrule
    CWE-022 (Path Traversal) & 48 & 31 & 64.6 & 0.75 \\
    CWE-079 (Cross-Site Scripting) & 36 & 18 & 50.0 & 0.621 \\
    CWE-094 (Code Injection) & 20 & 12 & 60.0 & 0.606 \\
    CWE-078 (OS Command Injection) & 12 & 7 & 58.3 & 0.628 \\
    CWE-502 (Deserialization) & 6 & 4 & 66.7 & 0.853 \\
    CWE-611 (XXE) & 5 & 3 & 60.0 & 0.657 \\
    Other CWEs ($\leq$4 CVEs) & 49 & 19 & 38.8 & 0.504 \\
    \midrule
    \textbf{Total} & \textbf{176} & \textbf{94} & \textbf{53.4} & \textbf{0.631} \\
    \bottomrule
    \end{tabular}
  \end{table}
  \begin{table}[htbp]
    \caption{Recall Performance Comparison Across Methods (Shared CVEs: 130).}
    \label{tab:method_comparison}
    \vspace{-10pt}
    \centering
    \scriptsize 
    \begin{tabular}{@{}lrrr@{}}
    \toprule
    \textbf{Method} & \textbf{Recall Rate (\%)} & \textbf{Avg Precision} & \textbf{Avg F1 Score} \\
    \midrule
    CodeQL & 20.0 & 0.055 & 0.073 \\
    IRIS & 35.4 & 0.031 & 0.048 \\
    \tool & 80.0 & 0.672 & 0.700 \\
    \bottomrule
    \end{tabular}
  \end{table}
\begin{figure}[t]
    \centering
    \includegraphics[width=0.9\linewidth]{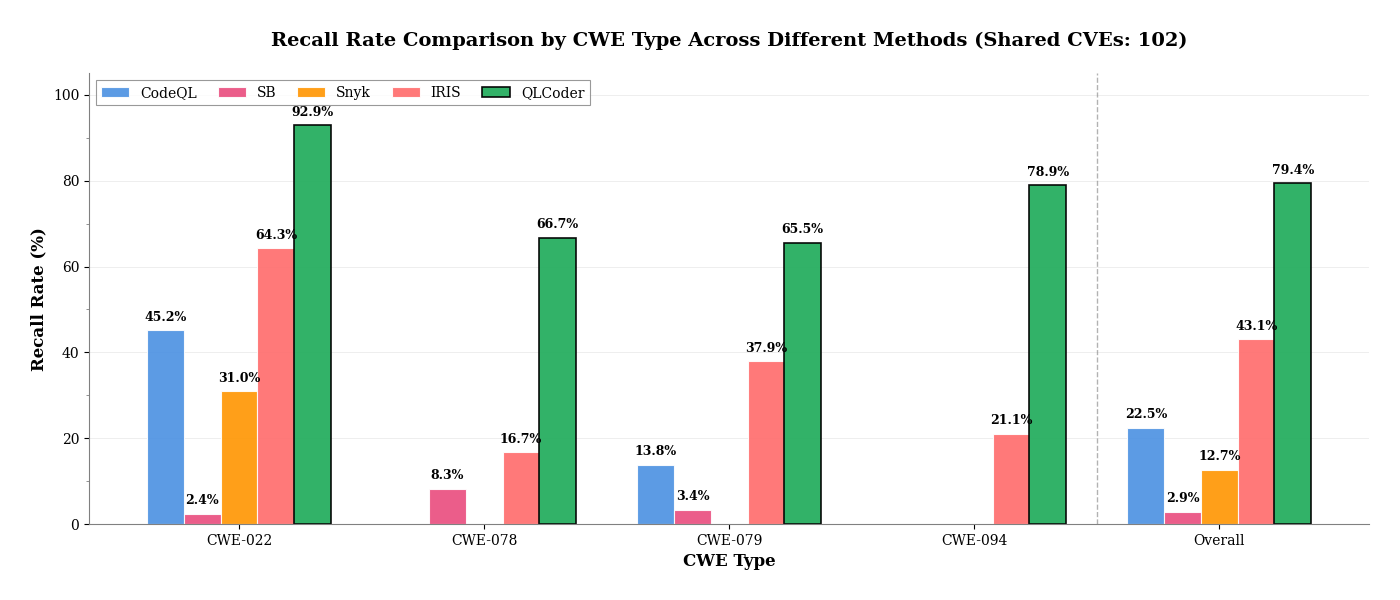}
    \vspace{-10px}
    \caption{Recall Rate Comparison by CWE Type Across Different Methods (102 CVEs).}
    \label{fig:sast_tool_comparison}
    \vspace{-0.1in}
\end{figure}
\paragraph{\tool vs.\ state-of-the-art QL.} 

Table~\ref{tab:cwe_success_results} shows \tool's overall query synthesis success rate by CWE. Table~\ref{tab:method_comparison} shows the notable increase in precision of \tool over CodeQL and IRIS. 
\tool is able to successfully synthesize 53.4\% of the CVEs. For half the queries \tool correctly synthesizes CodeQL, detects the CVE, and does not report false positives on the fixed version of the CVE's repository. The lack of true positive recall is why CodeQL and IRIS have significantly lower precision. CodeQL's queries are broad, categorized by CWE queries. IRIS generates all of the predicates for potential sources and sinks with CodeQL, and does not generate sanitizer or taint step predicates.  

Finally, Figure~\ref{fig:sast_tool_comparison} shows that CodeQL, IRIS, and \tool have significantly higher vulnerability recall rates compared to Snyk and SpotBugs. Thus highlighting CodeQL's superior performance compared to other static analysis tools.

\paragraph{Impact of training cut-off.} 
We also want to take note that Claude Sonnet 4's training cut-off is March 2025. Table~\ref{tab:performance_by_year} shows that \tool performs consistently regardless of CVEs before or after the cut-off period. The CodeQL version, 2.22.2, was released in July 2025. New versions of CodeQL often include analysis improvements and new QL packs \citep{noauthor_codeql_nodate-1}.

\begin{table}[htbp]
  \caption{Tool Performance Before vs After Training Cutoff.}
    \label{tab:performance_by_year}
    \vspace{-10pt}
    \centering
    \scriptsize 
    \begin{tabular}{@{}lrrrrr@{}}
    \toprule
    \textbf{CVE Period} & \textbf{Total CVEs} & \textbf{\# Recall} & \textbf{Success (\%)} & \textbf{Avg Precision} & \textbf{Avg F1 Score} \\
    \midrule
    Pre-2025 (2011-2024) & 111 & 64 & 57.7 & 0.676 & 0.702 \\
    2025+ (Post-cutoff) & 65 & 30 & 46.2 & 0.555 & 0.583 \\
    \midrule
    \textbf{Overall} & \textbf{176} & \textbf{94} & \textbf{53.4} & \textbf{0.631} & \textbf{0.658} \\
    \bottomrule
    \end{tabular}
  \end{table}


\begin{table}[htb]
\caption{Ablation Study (out of 20 CVEs).}
\centering
\scriptsize 
\vspace{-10pt}
  \label{tab:ablation_study}
  \begin{tabular}{lrrrr}
  \toprule
  \textbf{Variant} & \textbf{\% Successful} & \textbf{Recall Rate} & \textbf{Avg Precision} & \textbf{Avg F1 Score} \\
  \midrule
    \tool & \textbf{\datachange{50}\%} & \textbf{80\%} & \textbf{0.67} & \textbf{0.69} \\ \midrule
  w/o LSP & 25\% ($-$\datachange{25}\%) & 55\% ($-$25\%) & 0.32 & 0.36 \\
  w/o Doc/Ref & 20\% ($-$\datachange{30}\%) & 55\% ($-$25\%) & 0.32 & 0.36 \\
  w/o AST & 25\% ($-$\datachange{25}\%) & 80\% ($\pm$0\%) &  0.41 & 0.47 \\
  \changed{Claude Code} &  10\% ($-$\datachange{40}\%) & 55\% ($-$25\%) &  0.33 & 0.36 \\ 
  \bottomrule
  \end{tabular}
\end{table}
\vspace{-5px}
\subsection{RQ2: Ablation Studies}
\vspace{-.1in}
For ablations, we chose 20 CVEs and ran \tool with one of the \tool components removed~(Table~\ref{tab:ablation_study}). 
The ablation with no tools refers to only running Claude Code with the iterative feedback system. 
The high recall rate when removing access to the AST cache while lowered recall rates without the LSP server or documentation access show that the LSP and documentation lookup impact the synthesis performance more. 
We also include \tool's performance on the same set of CVEs, and point out its significantly higher query success rate and precision score. 
Claude Code without tools scored a high recall rate, yet failed to synthesize queries without false positives when executed on the fixed version.

\begin{table}[htbp]
\centering
\caption{LLM-agent baselines' compilation and success rates on 20 CVEs from 2025.}
\vspace{-10pt}
\label{tab:agent_baselines}
\scriptsize 
\begin{tabular}{llllll}
\toprule
\multirow{2}{*}{\textbf{Agent Baselines}}  & \multirow{2}{*}{\textbf{Language Model}} & 
\multicolumn{2}{c}{\textbf{w/o \tool}} & 
\multicolumn{2}{c}{\textbf{with \tool}} \\ \cmidrule{3-4} \cmidrule{5-6}
& & Compilation & Successful & Compilation & Successful \\
\midrule
Gemini CLI & Gemini 2.5 Flash & \datachange{0}\% & \changed{0\%} & \datachange{45}\% (+\datachange{45}\%) & \changed{0\%} (=) \\
Gemini CLI & Gemini 2.5 Pro & \changed{\datachange{60}\%} & \changed{0\%} & \changed{75\%} (+\datachange{15}\%) & \changed{0\%} (=) \\ \midrule
Codex & GPT-5, \changed{Minimal} & 0\% & \changed{0\%} & \datachange{35}\% (+\datachange{35}\%) & \changed{5\%} (+5\%) \\
\changed{Codex} & \changed{GPT-5, Medium} & \changed{0\%} & \changed{0\%} & \changed{\datachange{80}\%} (+\datachange{80}\%) & \changed{20\%} (+20\%) \\ 
\changed{Codex} & \changed{\texttt{gpt-oss-20b}} & \changed{25\%} & {0\%} & \changed{\datachange{45}\%} (+\datachange{20}\%) & \changed{5\%} (+5\%) \\ \midrule
\changed{Claude Code} & \changed{Sonnet 4} & \changed{95\%} & \changed{10\%} & \datachange{99\%} (+\datachange{4}\%) & \changed{35\%} (+25\%) \\
\bottomrule
\end{tabular}
\vspace{-0.15in}
\end{table}
\vspace{-5px}
\subsection{RQ3: State of the Art Agent Comparison}
\vspace{-.1in}
\changed{\tool can be transferred to other coding agents. For switching between coding agents, the MCP server configuration format may differ but the MCP server commands and arguments are the same. Coding agents also had different CLI commands used to invoke the agent. For the weaker, open source model we used (gpt-oss:20b), we had to break down our prompts into smaller tasks and increase query writing iterations.} We used Gemini CLI with Gemini 2.5 Flash \changed{and Gemini 2.5 Pro}. We used Codex with GPT-5 \changed{minimal reasoning effort, GPT-5 medium reasoning effort, and for trying open source models}. \changed{We used Codex with open source models due to ease of configuration. See Appendix \ref{sec:appendix-eval-detail} for more details on using Codex with open source models.} We evaluated their performance on 20 CVEs. We achieved an increase in compilation success for both agents compared to using the agents without \tool in Table~\ref{tab:agent_baselines}. \changed{We also achieved an increase in successful queries generated using Codex and GPT-5 medium reasoning with \tool. GPT-5 minimal reasoning with \tool also had a slight increase compared to not using, though not as large of a gain compared to using medium reasoning effort. Codex with gpt-oss-20b also had a slight increase in successful queries synthesized, compared to without \tool.} \changed{See Appendix \ref{sec:appendix-eval-detail} for statistics on common error categories for agent baseline failed queries.}
\vspace{-5px}
\changed{\subsection{RQ4: Scalability of \tool}}
\vspace{-.1in}

\changed{To compare costs and runtime of CodeQL query writing with \tool, do note that GitHub offers 8 consulting services related to CodeQL query writing to businesses \citep{github-services}. For \tool, the average time to synthesize each query took 3712 seconds. The average cost for each synthesized query was 2.90 USD. The average input token usage was 708 tokens per query. The average output token usage was 43176 tokens per query. Figure \ref{fig:sloc_duration_cost} shows how for queries that finish synthesizing in $<$ 3000 seconds, the success rate is around 97\%. Thus, one could consider an early stop policy around 3000 seconds. A more detailed breakdown of runtime and cost per CVE can be found in Appendix~\ref{sec:appendix-eval-detail}.}

\vspace{-20px}
\changed{\section{Discussion}}
 \vspace{-.3in}
\changed{\subsection{Versatility of \tool}}
 \vspace{-.1in}
\changed{\textbf{CodeQL Versions and Other Languages.~~} To upgrade the CodeQL version for \tool, the system has to fetch CodeQL documentation and queries related to the upgraded version. We used a simple python script to fetch CodeQL documentation web pages and each CodeQL CLI installation comes with its version-adjusted queries. To use \tool for other languages supported by CodeQL, the system has to fetch CodeQL documentation and queries specific to the language, which are categorized in the CLI installation and the site documentation.}

\changed{\textbf{Other SAST tools.~~} For adapting the system for other engines, Semgrep \citep{semgrep-lsp} and Snyk \citep{snyk-lsp} both have officially supported language servers. For engines that don't have language servers, one can create a MCP server that interfaces with calling the engine's commands. }

\vspace{-15px}
\changed{\subsection{Applications of \tool}}
\vspace{-.1in}
\changed{\textbf{Variant Analysis.~~} A \tool synthesized query can be used for variant analysis. By running the known CVE pattern on other repositories, we can find new bugs that match the pattern. CodeQL's VSCode extension already supports running a given query on the top 1000 public repositories or any specified group of repositories \citep{codeql_variant_analysis}. For example, with one \tool synthesized query we reported 2 unknown bugs in 2 different repositories.}

\changed{\textbf{Regression Testing.~~} In continuous integration pipelines, regression testing is common component to have. One of Github's CI features is its CodeQL code scan Github action \citep{code_scan}. The action automatically runs CodeQL queries on a given pull request or specified event. Currently CodeQL's provided security queries support only 64 CWEs \citep{noauthor_codeqljavaqlsrcsecuritycwe_nodate}. \tool is CWE agnostic and given a history of known CVEs for a specific repository, the synthesized queries can be used as regression testing via a CI pipeline.} 
\section{Related Work}
\vspace{-.1in}
\mypara{LLMs and vulnerability detection} LLMs have been used extensively for vulnerability detection and repair using techniques such as fine-tuning and prompt engineering \citep{zhou2024largelanguagemodelvulnerability}. LLMs have also been combined with existing program analysis tools for vulnerability detection. The combination of LLMs can be used from vulnerability analysis like IRIS's \citep{li2025iris} source and sink identification, however IRIS depends on a limited set of CWE templates derived from CodeQL's CWE queries. IRIS also only the LLM for identifying sources and sinks. KNighter synthesizes CSA checkers given a fix commit of a C repository \citep{yang2025knighter}, however the checkers are written in C which has more available training data. MocQ's uses an LLM to derive a subset DSL of CodeQL and Joern, and then provides a feedback loop to the LLM though prompting via API calls is used rather than an agent with tools and MocQ uses significantly higher iterations, with a max threshold of 1,000 iterations per vulnerability experiment. \citep{li2025automatedstaticvulnerabilitydetection}.

\mypara{LLM agents and tool usage} 
SWE-agent pioneered the idea of autonomous LLM agents using tools for software engineering tasks \cite{yang2024sweagentagentcomputerinterfacesenable}. LSPAI \cite{go_lspai_2025}, an IDE plugin, uses LSP servers to guide LLM-generated unit tests. Hazel, a live program sketching environment, uses a language server \citep{blinn_statically_2024} to assist code completions synthesized by LLMs. The Hazel Language Server provides the typing context of a program hole to be filled. 

\mypara{Low resource LLM code generation} SPEAC uses ASTs combined with constraint solving to repair LLM-generated code for low resource programming languages \citep{mora2024synthetic}. SPEAC converts a buggy program into an AST and uses a solver to find the minimum set of AST nodes to replace, to satisfy language constraints. MultiPL-T generates datasets for low resource languages by translating high resource language code to the target language and validates translations with LLM generated unit tests \citep{cassano2024knowledgetransferhighresourcelowresource}.
\section{Conclusion and Limitations} 
\vspace{-.1in}
We present \tool, an agentic framework for synthesizing syntactically correct and precise CodeQL queries given known vulnerability patterns. We will also open source our CodeQL LSP MCP server and \tool. 
In future work, we plan to explore efficient ways to synthesize, and to combine our synthesized queries with dynamic analysis tools. 

\textbf{Limitations.} 
We omit CVEs where the vulnerability involves non-Java code such as configuration files or other languages. 
\tool can be used with exploit generation to find vulnerabilities that are realized during dynamic execution. 
For supporting other languages that can be queried by CodeQL, the vector database can be filled with references, documentation, and example queries in other CodeQL supported languages. 
We also note that Claude Sonnet 4's official training cut-off is March 2025, however the 2025 CVEs evaluated were reported between January to August 2025. 

\section*{Acknowledgments} 

We thank the reviewers for feedback that improved this paper. \datachange{We thank Amartya Das for assisting with QLCoder open source efforts.} This research was supported by NSF award CCF \#2313010, NSF CISE Graduate Fellowships under Grant No. 2313998, and the Google TPU Builders Program. 

\bibliographystyle{iclr2025_conference}
\bibliography{iclr2025_conference}

\appendix
\section{Evaluation Details}\label{sec:appendix-eval-detail}
Table 6 is a more detailed breakdown of the successful query synthesis rate by CWE. \\
\\ 
\changed{\textbf{Runtime and Costs.} Table~\ref{tab:longtable} is a more detailed breakdown of cost and runtime per synthesized CVE query. In table 7, SLOC is the number of source lines of code. SLOC is calculated by using the CLI tool tokei \citep{tokei} on the project checked out at the patch commit and counting the number of lines of code for Java files of the project. Time was derived from recording the timestamps of starting \tool and when \tool returned a successful query or exhausted 10 iterations of the query. Note that the analysis phase invokes Claude Code once, and the query writing phase invokes Claude Code up to 10 times. Thus for deriving total tokens used, we summed up the following token values found from the following keys in each Claude Code API response (as of September 24, 2025) - \texttt{input\_tokens}, \texttt{cache\_creation\_input\_tokens}, \texttt{cache\_read\_input\_tokens}, and \texttt{output\_tokens}. In Table~\ref{tab:longtable}, the Tokens column shows token counts in thousands. For the USD cost value, we added up each Claude Code API response's \texttt{total\_cost\_usd} value in a given CVE synthesis. In Table~\ref{tab:longtable}, the USD column shows the total cost of synthesizing the query.}

\begin{figure}[t]
    \centering
    \includegraphics[width=0.9\linewidth]{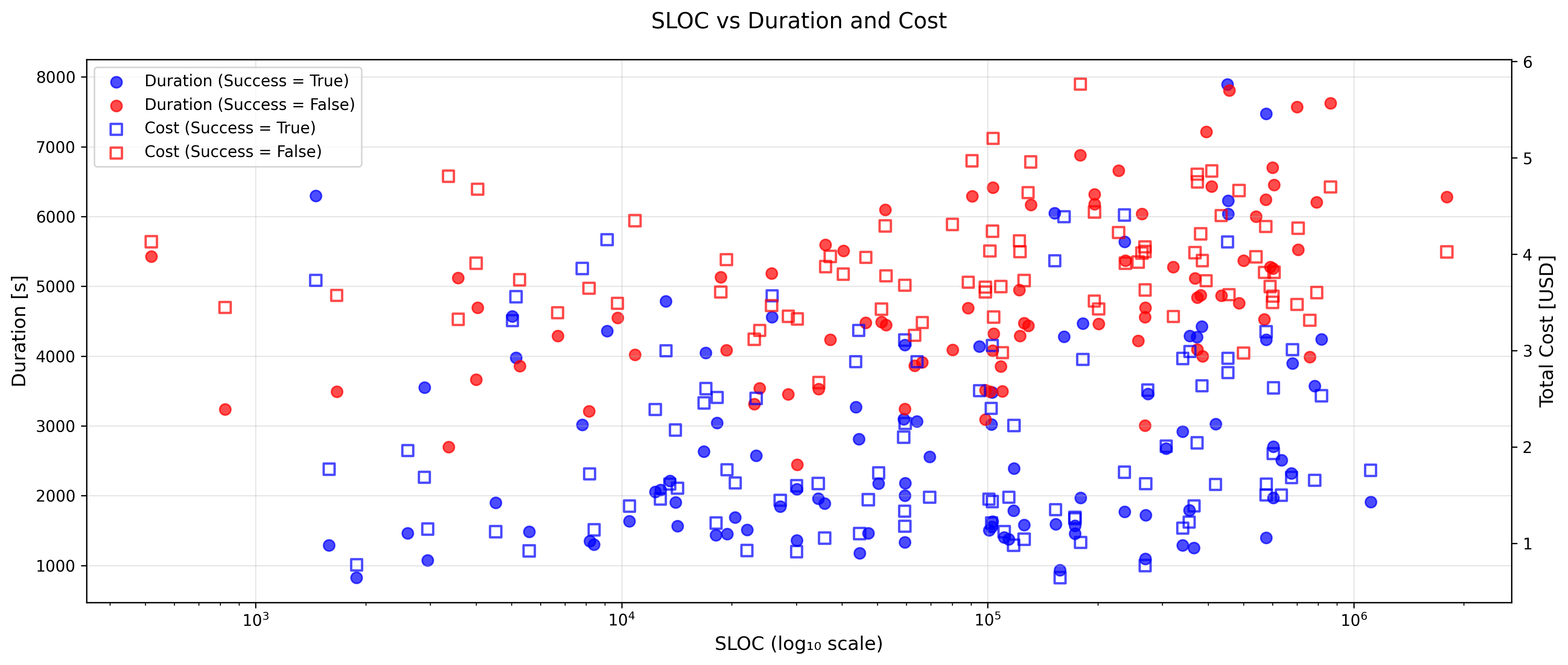}
    \vspace{-10px}
    \caption{SLOC, Synthesis Duration, and Synthesis Cost Comparison by CVE.}
    \label{fig:sloc_duration_cost}
    \vspace{-0.1in}
\end{figure}

\begin{figure}[t]
    \centering
    \includegraphics[width=\linewidth]{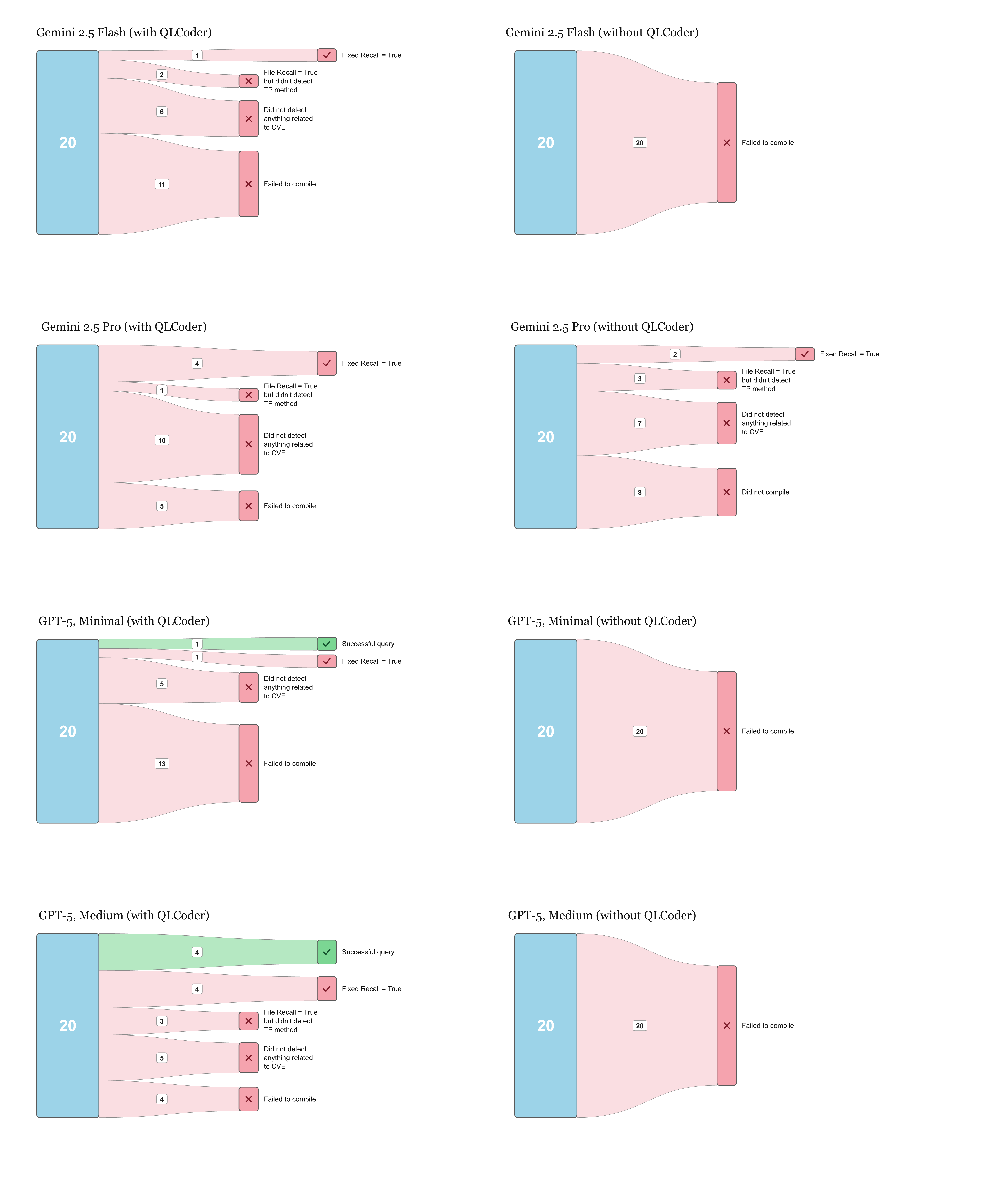}
    \caption{
        \changed{\datachange{Error Decomposition for Table} \ref{tab:agent_baselines}.}
    }
    \label{fig:agent-error-categories}
    \vspace{-0.15in}
\end{figure}

\changed{\textbf{Error Categories for Agent Baselines' from Table \ref{tab:agent_baselines}.} We derived the error categories for why a query wasn't successful in each of the agent baseline configurations. The categories are: (i) the patched version has vulnerability recall, (ii) the query did not detect any of the patched methods related to the CVE, (iii) the query did have a file recall but failed to alarm on the vulnerable methods in the file, and (iv) the query did not compile. We have also included Figure \ref{fig:agent-error-categories} to visualize each agent baseline configurations' error category count.}

\changed{\textbf{CodeQL Baseline Queries.~~} The CodeQL baseline queries used are from CodeQL's security QL pack \citep{noauthor_codeqljavaqlsrcsecuritycwe_nodate}. We picked the CVEs that have CWEs supported by CodeQL's security queries. Not all of CWEs represented in our evaluation have corresponding CodeQL CWE queries.}

\changed{\textbf{Using Codex with Open Source Models.~~} As of November 17, 2025, to use Codex with open source models, we used ollama \citep{ollama} to download and host the models. Codex has a configuration file located at \texttt{[home directory]/.codex/config.toml}. For each open source model we wanted to try out with \tool, we added a new profile for the model in the configuration file. We also were able to use existing MCP server configurations for Codex, with the open source models. Below is an example of the configuration for using gpt-oss:20b with Codex. Afterwards, to use the profile with Codex we use the command \texttt{codex --profile gpt-oss-20b}}.

\begin{lstlisting}[language=toml,frame=single]
[model_providers.ollama]
name = "Ollama"
base_url = "http://localhost:11434/v1"
[profiles.gpt-oss-20b]
model_provider = "ollama"
model = "gpt-oss:20b"
wire_api = "chat"
\end{lstlisting}

\subsection{Evaluation Limitations}\label{sec:appendix-evaluation-limit}
\textbf{Codex CLI.} Claude Code and Gemini CLI allow users to configure how many max turns an agent can take in a context window. As of 9/23/2025, Codex CLI does not offer this configuration. Thus we were not able to force Codex to always take up to 50 max turns each context window. \\
\\
\textbf{IRIS.} The original IRIS evaluation consists of 120 Java projects from CWE-Bench-Java. Many of these projects are old with deprecated dependencies, thus we were only able to build and use 112 of the projects with CodeQL 2.22.2. As of 9/23/2025, IRIS supports 11 CWEs and out of the 65 CVEs from 2025, we were able to use 24 of them with IRIS. When running some of the IRIS queries, the amount of sources and sink predicates in the query led to out of memory errors. This impacted 9 out of the 24 queries, thus we treat those as queries with 0 results and false recall.   \\

\begin{table}[htbp]
    \caption{\tool Query Success by CWE Type.}
    \label{tab:cwe_detection_results}
    \centering
    \begin{tabular}{@{}l@{\hspace{0.5em}}rrrr@{}}
    \toprule
    \textbf{CWE Type} & \textbf{Total CVEs} & \textbf{\# Success} & \textbf{Success (\%)} & \textbf{Avg Precision} \\
    \midrule
    CWE-022 (Path Traversal) & 48 & 31 & 64.6\% & 0.750 \\
    CWE-079 (Cross-Site Scripting) & 36 & 18 & 50.0\% & 0.621 \\
    CWE-094 (Code Injection) & 20 & 12 & 60.0\% & 0.606 \\
    CWE-078 (OS Command Injection) & 12 & 7 & 58.3\% & 0.628 \\
    CWE-502 (Deserialization) & 6 & 4 & 66.7\% & 0.853 \\
    CWE-611 (XXE) & 5 & 3 & 60.0\% & 0.657 \\
    CWE-287 (Authentication) & 4 & 1 & 25.0\% & 0.875 \\
    CWE-200 (Information Exposure) & 3 & 0 & 0.0\% & 0.667 \\
    CWE-400 (Resource Consumption) & 3 & 1 & 33.3\% & 0.556 \\
    CWE-532 (Information Exposure) & 3 & 3 & 100.0\% & 0.686 \\
    CWE-770 (Resource Exhaustion) & 3 & 1 & 33.3\% & 0.444 \\
    CWE-020 (Improper Input Validation) & 2 & 2 & 100.0\% & 0.650 \\
    CWE-089 (SQL Injection) & 2 & 2 & 100.0\% & 1.000 \\
    CWE-1333 (ReDoS) & 2 & 0 & 0.0\% & 0.000 \\
    CWE-284 (Access Control) & 2 & 0 & 0.0\% & 0.500 \\
    CWE-862 (Authorization) & 2 & 0 & 0.0\% & 0.000 \\
    CWE-918 (SSRF) & 2 & 1 & 50.0\% & 0.500 \\
    CWE-023 (Relative Path Traversal) & 1 & 1 & 100.0\% & 1.000 \\
    CWE-044 (Path Equivalence) & 1 & 1 & 100.0\% & 0.667 \\
    CWE-083 (Improper Neutralization) & 1 & 1 & 100.0\% & 0.052 \\
    CWE-1325 (Improperly Controlled Memory) & 1 & 0 & 0.0\% & 0.000 \\
    CWE-164 (Foreign Code) & 1 & 0 & 0.0\% & 0.000 \\
    CWE-178 (Case Sensitivity) & 1 & 0 & 0.0\% & 1.000 \\
    CWE-190 (Integer Overflow) & 1 & 0 & 0.0\% & 0.000 \\
    CWE-264 (Permissions) & 1 & 0 & 0.0\% & 0.000 \\
    CWE-267 (Privilege Defined) & 1 & 0 & 0.0\% & 0.000 \\
    CWE-276 (Incorrect Permissions) & 1 & 0 & 0.0\% & 1.000 \\
    CWE-285 (Improper Authorization) & 1 & 1 & 100.0\% & 1.000 \\
    CWE-288 (Authentication Bypass) & 1 & 0 & 0.0\% & 0.000 \\
    CWE-290 (Authentication Bypass) & 1 & 1 & 100.0\% & 1.000 \\
    CWE-297 (Improper Certificate) & 1 & 1 & 100.0\% & 1.000 \\
    CWE-312 (Cleartext Storage) & 1 & 0 & 0.0\% & 0.000 \\
    CWE-327 (Cryptographic Issues) & 1 & 0 & 0.0\% & 0.000 \\
    CWE-346 (Origin Validation) & 1 & 0 & 0.0\% & 0.200 \\
    CWE-352 (CSRF) & 1 & 1 & 100.0\% & 0.941 \\
    CWE-426 (Untrusted Search Path) & 1 & 0 & 0.0\% & 0.000 \\
    CWE-835 (Infinite Loop) & 1 & 0 & 0.0\% & 0.000 \\
    CWE-863 (Authorization) & 1 & 1 & 100.0\% & 1.000 \\
    \midrule
    \textbf{Total} & \textbf{176} & \textbf{94} & \textbf{53.4\%} & \textbf{0.631} \\
    \bottomrule
    \end{tabular}
  \end{table}

\changed{
\setlength{\tabcolsep}{1pt}
\begin{longtable}{llllrrr}
\caption{\changed{\tool duration, costs, and total tokens used per CVE ID.}}
\label{tab:longtable} \\ \toprule
\textbf{CVE-ID} & \textbf{CWE-ID} & \textbf{Project} & \textbf{SLOC} & \textbf{Time (s)} & \textbf{Tokens (K)} & \textbf{USD} \\ \midrule
\endfirsthead
\toprule
\textbf{CVE-ID} & \textbf{CWE-ID} & \textbf{Project} & \textbf{SLOC} & \textbf{Time (s)} & \textbf{Tokens (K)} & \textbf{USD} \\ \midrule
\endhead
\midrule
\multicolumn{7}{r}{\textit{Continued on next page}} \\
\endfoot
\midrule
\multicolumn{3}{l}{\textbf{Average}} & \textbf{209.44K} & \textbf{3712} & \textbf{4614.56} & \textbf{2.90} \\
\bottomrule
\endlastfoot
    CVE-2025-24970 & 020 & netty & 341.20K & 2916 & 5060.08 & 2.92 \\
    CVE-2025-22233 & 020 & spring-framework & 782.76K & 3570 & 2750.30 & 1.65 \\
    CVE-2022-23082 & 022 & CureKit & 519 & 5429 & 6415.35 & 4.13 \\
    CVE-2022-31194 & 022 & DSpace & 237.97K & 5367 & 5934.22 & 3.91 \\
    CVE-2022-31195 & 022 & DSpace & 236.54K & 5636 & 5346.47 & 4.41 \\
    CVE-2025-53622 & 022 & DSpace & 373.20K & 4275 & 3052.49 & 2.04 \\
    CVE-2018-12036 & 022 & DependencyCheck & 28.53K & 3453 & 5467.16 & 3.36 \\
    CVE-2021-29425 & 022 & commons-io & 40.33K & 5511 & 5844.67 & 3.79 \\
    CVE-2025-0851 & 022 & djl & 102.97K & 3477 & 4968.76 & 3.05 \\
    CVE-2022-26884 & 022 & dolphinscheduler & 90.69K & 6288 & 7767.97 & 4.97 \\
    CVE-2022-34662 & 022 & dolphinscheduler & 110.91K & 1402 & 1748.13 & 1.12 \\
    CVE-2022-23457 & 022 & esapi-java-legacy & 35.88K & 1888 & 1529.97 & 1.05 \\
    CVE-2020-17519 & 022 & flink & 1.11M & 1912 & 3395.07 & 1.76 \\
    CVE-2022-26049 & 022 & goomph & 12.77K & 2084 & 2150.50 & 1.46 \\
    CVE-2023-41044 & 022 & graylog2-server & 270.29K & 1722 & 2778.67 & 1.62 \\
    CVE-2018-17297 & 022 & hutool & 64.12K & 3063 & 4408.37 & 2.88 \\
    CVE-2022-22931 & 022 & james-project & 434.53K & 4865 & 6468.51 & 4.40 \\
    CVE-2020-8570 & 022 & java & 816.52K & 4238 & 3653.89 & 2.53 \\
    CVE-2025-49656 & 022 & jena & 676.42K & 2319 & 2795.36 & 1.68 \\
    CVE-2019-0225 & 022 & jspwiki & 59.35K & 2002 & 2183.08 & 1.33 \\
    CVE-2022-3782 & 022 & keycloak & 576.48K & 4234 & 2193.97 & 1.50 \\
    CVE-2023-35887 & 022 & mina-sshd & 131.18K & 6165 & 7486.34 & 4.96 \\
    CVE-2020-35460 & 022 & mpxj & 181.97K & 4468 & 4041.71 & 2.91 \\
    CVE-2011-4367 & 022 & myfaces & 263.92K & 6037 & 6228.39 & 4.01 \\
    CVE-2022-25842 & 022 & one-java-agent & 5.59K & 1481 & 1118.59 & 0.92 \\
    CVE-2018-1002200 & 022 & plexus-archiver & 13.54K & 2213 & 2585.96 & 1.61 \\
    CVE-2023-37460 & 022 & plexus-archiver & 12.36K & 2056 & 3958.90 & 2.39 \\
    CVE-2022-4244 & 022 & plexus-utils & 23.84K & 3540 & 5521.94 & 3.21 \\
    CVE-2018-1000850 & 022 & retrofit & 20.43K & 1688 & 2831.06 & 1.63 \\
    CVE-2019-17572 & 022 & rocketmq & 94.98K & 4136 & 3845.71 & 2.58 \\
    CVE-2023-34478 & 022 & shiro & 37.14K & 4233 & 6733.75 & 3.98 \\
    CVE-2023-46749 & 022 & shiro & 43.62K & 3268 & 4812.75 & 2.89 \\
    CVE-2024-23673 & 022 & sling-org-apache-sling-servlets-resolver & 8.41K & 1297 & 1730.85 & 1.14 \\
    CVE-2025-46096 & 022 & solon & 125.80K & 1579 & 1695.92 & 1.04 \\
    CVE-2016-9177 & 022 & spark & 9.74K & 4547 & 5257.03 & 3.49 \\
    CVE-2018-9159 & 022 & spark & 10.87K & 4017 & 6149.27 & 4.35 \\
    CVE-2020-5405 & 022 & spring-cloud-config & 22.01K & 1511 & 1414.43 & 0.92 \\
    CVE-2020-5410 & 022 & spring-cloud-config & 18.66K & 5131 & 5543.48 & 3.61 \\
    CVE-2019-0207 & 022 & tapestry-5 & 157.63K & 932 & 1092.47 & 0.64 \\
    CVE-2018-11762 & 022 & tika & 109.76K & 3494 & 4456.12 & 2.98 \\
    CVE-2022-32287 & 022 & uima-uimaj & 227.86K & 6658 & 5932.88 & 4.23 \\
    CVE-2014-7816 & 022 & undertow & 103.90K & 4320 & 5419.59 & 3.35 \\
    CVE-2022-36007 & 022 & venice & 117.70K & 1783 & 1371.73 & 0.98 \\
    CVE-2018-12542 & 022 & vertx-web & 34.46K & 1961 & 2487.06 & 1.62 \\
    CVE-2019-17640 & 022 & vertx-web & 47.21K & 1462 & 2660.48 & 1.45 \\
    CVE-2018-1047 & 022 & wildfly & 486.38K & 4759 & 7287.55 & 4.66 \\
    CVE-2023-45277 & 022 & yamcs & 173.18K & 1457 & 2031.62 & 1.27 \\
    CVE-2023-45278 & 022 & yamcs & 173.18K & 1572 & 2204.78 & 1.25 \\
    CVE-2018-1002202 & 022 & zip4j & 8.18K & 1348 & 2894.83 & 1.72 \\
    CVE-2018-1002201 & 022 & zt-zip & 7.80K & 3016 & 5901.48 & 3.85 \\
    CVE-2025-1584 & 023 & solon & 118.06K & 2388 & 2999.88 & 2.22 \\
    CVE-2025-24813 & 044 & tomcat & 366.29K & 1252 & 2409.97 & 1.39 \\
    CVE-2025-26074 & 078 & conductor & 114.24K & 1376 & 2726.99 & 1.48 \\
    CVE-2022-20617 & 078 & docker-commons-plugin & 2.89K & 3551 & 2354.26 & 1.68 \\
    CVE-2019-10392 & 078 & git-client-plugin & 16.79K & 2630 & 4160.82 & 2.46 \\
    CVE-2020-2261 & 078 & perfecto-plugin & 825 & 3238 & 5544.05 & 3.45 \\
    CVE-2017-1000487 & 078 & plexus-utils & 23.30K & 2571 & 3938.81 & 2.50 \\
    CVE-2023-24422 & 078 & script-security-plugin & 8.15K & 3211 & 6250.12 & 3.65 \\
    CVE-2022-25174 & 078 & workflow-cps-global-lib-plugin & 4.53K & 1901 & 1634.79 & 1.12 \\
    CVE-2022-25173 & 078 & workflow-cps-plugin & 16.98K & 4048 & 3594.86 & 2.61 \\
    CVE-2022-25175 & 078 & workflow-multibranch-plugin & 3.58K & 5120 & 4783.47 & 3.32 \\
    CVE-2013-7285 & 078 & xstream & 44.44K & 2809 & 5416.36 & 3.21 \\
    CVE-2020-26217 & 078 & xstream & 52.51K & 6096 & 6573.86 & 4.29 \\
    CVE-2021-21345 & 078 & xstream & 52.77K & 4443 & 5313.38 & 3.78 \\
    CVE-2022-31192 & 079 & DSpace & 236.52K & 1767 & 2638.92 & 1.74 \\
    CVE-2016-10006 & 079 & antisamy & 4.01K & 3664 & 6565.33 & 3.91 \\
    CVE-2017-14735 & 079 & antisamy & 4.04K & 4696 & 7181.97 & 4.68 \\
    CVE-2022-28367 & 079 & antisamy & 5.03K & 4569 & 4505.33 & 3.31 \\
    CVE-2022-29577 & 079 & antisamy & 5.14K & 3973 & 5344.35 & 3.56 \\
    CVE-2025-47885 & 079 & cloudbees-jenkins-advisor-plugin & 2.60K & 1461 & 3818.47 & 1.96 \\
    CVE-2025-2901 & 079 & console & 122.59K & 4287 & 5209.18 & 4.03 \\
    CVE-2016-6812 & 079 & cxf & 602.83K & 2700 & 2878.97 & 1.93 \\
    CVE-2019-17573 & 079 & cxf & 680.23K & 3897 & 4745.86 & 3.01 \\
    CVE-2022-24891 & 079 & esapi-java-legacy & 36.07K & 5593 & 5719.59 & 3.87 \\
    CVE-2020-27219 & 079 & hawkbit & 103.46K & 6416 & 6996.26 & 5.20 \\
    CVE-2019-10219 & 079 & hibernate-validator & 88.47K & 4691 & 6373.29 & 3.71 \\
    CVE-2018-1000129 & 079 & jolokia & 30.21K & 2441 & 5901.65 & 3.33 \\
    CVE-2025-32961 & 079 & jpawebapi & 3.37K & 2698 & 7642.30 & 4.81 \\
    CVE-2020-13973 & 079 & json-sanitizer & 1.67K & 3489 & 5929.25 & 3.57 \\
    CVE-2019-10076 & 079 & jspwiki & 59.38K & 3244 & 5633.10 & 3.68 \\
    CVE-2019-10077 & 079 & jspwiki & 59.38K & 1335 & 1950.83 & 1.18 \\
    CVE-2019-10078 & 079 & jspwiki & 59.38K & 4160 & 4980.42 & 3.11 \\
    CVE-2019-10089 & 079 & jspwiki & 59.55K & 2177 & 3361.70 & 2.24 \\
    CVE-2022-46907 & 079 & jspwiki & 58.95K & 3097 & 3359.56 & 2.10 \\
    CVE-2023-33962 & 079 & jstachio & 19.33K & 4083 & 7190.32 & 3.94 \\
    CVE-2025-23026 & 079 & jte & 23.05K & 3313 & 5277.52 & 3.12 \\
    CVE-2014-3656 & 079 & keycloak & 66.27K & 3912 & 5476.51 & 3.29 \\
    CVE-2022-1274 & 079 & keycloak & 576.88K & 7472 & 4463.84 & 3.20 \\
    CVE-2022-4137 & 079 & keycloak & 576.99K & 1397 & 2976.16 & 1.61 \\
    CVE-2022-4361 & 079 & keycloak & 575.34K & 6241 & 7337.30 & 4.29 \\
    CVE-2021-44667 & 079 & nacos & 122.11K & 4944 & 5964.66 & 4.14 \\
    CVE-2025-32960 & 079 & restapi & 19.41K & 1452 & 3001.09 & 1.76 \\
    CVE-2016-5394 & 079 & sling-org-apache-sling-xss & 1.46K & 6294 & 5968.91 & 3.73 \\
    CVE-2025-46558 & 079 & syntax-markdown & 2.95K & 1075 & 2051.41 & 1.15 \\
    CVE-2023-29201 & 079 & xwiki-commons & 101.47K & 3493 & 7209.06 & 4.03 \\
    CVE-2023-29528 & 079 & xwiki-commons & 102.52K & 1555 & 2139.98 & 1.21 \\
    CVE-2023-31126 & 079 & xwiki-commons & 103.00K & 1627 & 2454.59 & 1.43 \\
    CVE-2023-36471 & 079 & xwiki-commons & 103.17K & 4079 & 6791.06 & 4.24 \\
    CVE-2023-37908 & 079 & xwiki-rendering & 51.28K & 4486 & 5594.74 & 3.43 \\
    CVE-2020-29204 & 079 & xxl-job & 9.12K & 4361 & 6101.32 & 4.15 \\
    CVE-2023-32070 & 083 & xwiki-rendering & 50.36K & 2175 & 2917.20 & 1.73 \\
    CVE-2022-45206 & 089 & jeecgboot & 100.87K & 1505 & 2707.01 & 1.46 \\
    CVE-2022-4963 & 089 & spring-module-core & 1.59K & 1287 & 3680.27 & 1.77 \\
    CVE-2019-0222 & 094 & activemq & 409.42K & 6429 & 7499.98 & 4.86 \\
    CVE-2020-11998 & 094 & activemq & 419.39K & 3025 & 2437.65 & 1.61 \\
    CVE-2022-42889 & 094 & commons-text & 27.12K & 1842 & 2152.16 & 1.45 \\
    CVE-2021-41269 & 094 & cron-utils & 13.21K & 4784 & 4689.80 & 3.00 \\
    CVE-2023-49109 & 094 & dolphinscheduler & 153.39K & 1591 & 2369.87 & 1.35 \\
    CVE-2023-51770 & 094 & dolphinscheduler & 152.57K & 6049 & 6345.60 & 3.93 \\
    CVE-2021-30180 & 094 & dubbo & 179.47K & 1968 & 1735.59 & 1.01 \\
    CVE-2021-30181 & 094 & dubbo & 179.12K & 6880 & 8732.56 & 5.76 \\
    CVE-2022-44262 & 094 & ff4j & 46.47K & 4480 & 6612.80 & 3.97 \\
    CVE-2021-4178 & 094 & kubernetes-client & 69.41K & 2556 & 2116.29 & 1.48 \\
    CVE-2025-30067 & 094 & kylin & 395.31K & 7213 & 5382.61 & 3.72 \\
    CVE-2023-34468 & 094 & nifi & 864.61K & 7622 & 8168.47 & 4.70 \\
    CVE-2023-33246 & 094 & rocketmq & 108.61K & 3850 & 5853.05 & 3.66 \\
    CVE-2023-37582 & 094 & rocketmq & 201.07K & 4461 & 5656.22 & 3.43 \\
    CVE-2022-46166 & 094 & spring-boot-admin & 18.23K & 3040 & 4401.71 & 2.51 \\
    CVE-2022-22947 & 094 & spring-cloud-gateway & 25.77K & 4561 & 3338.67 & 3.57 \\
    CVE-2022-22965 & 094 & spring-framework & 705.22K & 5523 & 6182.90 & 4.27 \\
    CVE-2018-1260 & 094 & spring-security-oauth & 44.68K & 1176 & 1851.95 & 1.10 \\
    CVE-2023-32697 & 094 & sqlite-jdbc & 18.11K & 1436 & 2001.01 & 1.21 \\
    CVE-2020-17530 & 094 & struts & 161.73K & 4277 & 7378.37 & 4.39 \\
    CVE-2025-2240 & 1325 & smallrye-fault-tolerance & 34.56K & 3527 & 4291.18 & 2.67 \\
    CVE-2025-48058 & 1333 & powsybl-core & 269.01K & 4560 & 6698.58 & 4.08 \\
    CVE-2025-48059 & 1333 & powsybl-core & 269.35K & 4691 & 6438.45 & 4.03 \\
    CVE-2025-30177 & 164 & camel & 1.80M & 6280 & 7088.64 & 4.02 \\
    CVE-2025-24399 & 178 & oic-auth-plugin & 5.26K & 3860 & 5949.46 & 3.73 \\
    CVE-2025-52520 & 190 & tomcat & 381.92K & 4871 & 6579.21 & 4.21 \\
    CVE-2025-26795 & 200 & iotdb & 759.13K & 3989 & 5053.95 & 3.31 \\
    CVE-2025-54380 & 200 & opencast & 257.74K & 4220 & 6120.22 & 3.92 \\
    CVE-2025-22227 & 200 & reactor-netty & 80.17K & 4088 & 7084.78 & 4.31 \\
    CVE-2014-3576 & 264 & activemq & 321.48K & 5276 & 5521.34 & 3.35 \\
    CVE-2025-23015 & 267 & cassandra & 500.47K & 5368 & 4546.67 & 2.98 \\
    CVE-2025-24790 & 276 & snowflake-jdbc & 98.59K & 3512 & 5499.07 & 3.66 \\
    CVE-2025-48734 & 284 & commons-beanutils & 25.69K & 5183 & 5848.92 & 3.47 \\
    CVE-2025-1391 & 284 & keycloak & 591.27K & 5277 & 6025.32 & 3.67 \\
    CVE-2025-53106 & 285 & graylog2-server & 356.87K & 4287 & 5071.88 & 2.99 \\
    CVE-2023-51982 & 287 & crate & 540.71K & 5998 & 6567.43 & 3.97 \\
    CVE-2025-0604 & 287 & keycloak & 570.32K & 4526 & 6785.14 & 3.81 \\
    CVE-2025-3910 & 287 & keycloak & 600.36K & 6701 & 5630.84 & 3.50 \\
    CVE-2025-22228 & 287 & spring-security & 273.92K & 3460 & 4215.36 & 2.59 \\
    CVE-2025-49125 & 288 & tomcat & 374.12K & 4096 & 7034.93 & 4.75 \\
    CVE-2025-22223 & 290 & spring-security & 307.39K & 2675 & 3320.81 & 2.01 \\
    CVE-2025-3501 & 297 & keycloak & 603.77K & 1971 & 4672.52 & 2.61 \\
    CVE-2025-53103 & 312 & junit-framework & 125.80K & 4471 & 5648.19 & 3.73 \\
    CVE-2025-27508 & 327 & emissary & 63.19K & 3865 & 5184.05 & 3.16 \\
    CVE-2025-7365 & 346 & keycloak & 606.26K & 6454 & 5955.44 & 3.81 \\
    CVE-2025-31723 & 352 & simple-queue-plugin & 1.89K & 824 & 1248.24 & 0.78 \\
    CVE-2025-48795 & 400 & cxf & 700.41K & 7568 & 5211.09 & 3.48 \\
    CVE-2025-25193 & 400 & netty & 341.19K & 1290 & 1896.56 & 1.16 \\
    CVE-2025-53506 & 400 & tomcat & 373.95K & 4838 & 7602.65 & 4.83 \\
    CVE-2025-24789 & 426 & snowflake-jdbc & 98.59K & 3089 & 5808.05 & 3.61 \\
    CVE-2025-27522 & 502 & inlong & 456.66K & 7805 & 5149.58 & 3.58 \\
    CVE-2025-27526 & 502 & inlong & 453.50K & 6227 & 4254.62 & 2.77 \\
    CVE-2025-27528 & 502 & inlong & 453.50K & 6036 & 4720.41 & 2.92 \\
    CVE-2025-27531 & 502 & inlong & 452.17K & 7894 & 6402.01 & 4.13 \\
    CVE-2025-27818 & 502 & kafka & 792.42K & 6204 & 5508.74 & 3.60 \\
    CVE-2025-47771 & 502 & powsybl-core & 269.40K & 1094 & 1366.41 & 0.77 \\
    CVE-2025-48955 & 532 & para & 30.08K & 2094 & 2722.48 & 1.60 \\
    CVE-2025-49009 & 532 & para & 30.08K & 1357 & 1624.77 & 0.91 \\
    CVE-2025-27496 & 532 & snowflake-jdbc & 102.30K & 3024 & 4348.54 & 2.40 \\
    CVE-2025-53621 & 611 & DSpace & 386.12K & 3998 & 6407.57 & 3.93 \\
    CVE-2025-52888 & 611 & allure2 & 14.21K & 1564 & 2836.58 & 1.57 \\
    CVE-2025-27136 & 611 & local-s3 & 14.03K & 1904 & 3533.41 & 2.18 \\
    CVE-2025-47293 & 611 & powsybl-core & 269.25K & 3006 & 6040.41 & 3.63 \\
    CVE-2025-4641 & 611 & webdrivermanager & 10.50K & 1633 & 2034.01 & 1.39 \\
    CVE-2025-48976 & 770 & commons-fileupload & 6.69K & 4289 & 5690.52 & 3.39 \\
    CVE-2025-32959 & 770 & cuba & 384.90K & 4424 & 4115.30 & 2.63 \\
    CVE-2025-48988 & 770 & tomcat & 368.45K & 5115 & 6289.47 & 4.02 \\
    CVE-2025-27497 & 835 & OpenDJ & 601.66K & 5256 & 5809.20 & 3.56 \\
    CVE-2025-31720 & 862 & jenkins & 195.74K & 6319 & 5288.47 & 3.51 \\
    CVE-2025-31721 & 862 & jenkins & 195.81K & 6179 & 7173.86 & 4.44 \\
    CVE-2025-27427 & 863 & activemq-artemis & 634.70K & 2506 & 2809.01 & 1.50 \\
    CVE-2023-51441 & 918 & axis-axis1-java & 128.99K & 4432 & 8246.73 & 4.64 \\
    CVE-2024-48944 & 918 & kylin & 355.34K & 1791 & 1932.82 & 1.22 \\
\end{longtable}%
}

\setlength{\tabcolsep}{6pt}
\begin{longtable}{lll}
\caption{\changed{\datachange{List of the 20 CVEs for the ablation studies from Table} \ref{tab:ablation_study}.}}
\label{tab:longtable} \\ \toprule
\textbf{CVE-ID} & \textbf{CWE-ID} & \textbf{Project} \\ \midrule
\endfirsthead
\toprule
\textbf{CVE-ID} & \textbf{CWE-ID} & \textbf{Project} \\ \midrule
\endhead
\midrule
\multicolumn{3}{r}{\textit{Continued on next page}} \\
\endfoot
\bottomrule
\endlastfoot
    CVE-2025-52520 & 190 & tomcat \\
    CVE-2025-22227 & 200 & reactor-netty \\
    CVE-2025-49656 & 022 & jena \\
    CVE-2025-53622 & 022 & DSpace \\
    CVE-2025-1584 & 023 & solon \\
    CVE-2025-1391 & 284 & keycloak \\
    CVE-2025-48734 & 284 & commons-beanutils \\
    CVE-2025-0604 & 287 & keycloak \\
    CVE-2025-22223 & 290 & spring-security \\
    CVE-2025-53103 & 312 & junit-framework \\
    CVE-2025-53506 & 400 & tomcat \\
    CVE-2025-47771 & 502 & powsybl-core \\
    CVE-2025-48955 & 532 & para \\
    CVE-2025-49009 & 532 & para \\
    CVE-2025-4641 & 611 & webdrivermanager \\
    CVE-2025-52888 & 611 & allure2 \\
    CVE-2025-53621 & 611 & DSpace \\
    CVE-2025-48976 & 770 & commons-fileupload \\
    CVE-2025-48988 & 770 & tomcat \\
    CVE-2025-46558 & 079 & syntax-markdown \\
\end{longtable}%

\setlength{\tabcolsep}{6pt}
\begin{longtable}{lll}
\caption{\changed{\datachange{List of the 20 CVEs for the cross model evaluation from Table} \ref{tab:agent_baselines}.}}
\label{tab:longtable} \\ \toprule
\textbf{CVE-ID} & \textbf{CWE-ID} & \textbf{Project} \\ \midrule
\endfirsthead
\toprule
\textbf{CVE-ID} & \textbf{CWE-ID} & \textbf{Project} \\ \midrule
\endhead
\midrule
\multicolumn{3}{r}{\textit{Continued on next page}} \\
\endfoot
\bottomrule
\endlastfoot
    CVE-2025-24399 & 178 & oic-auth-plugin \\
    CVE-2025-22227 & 200 & reactor-netty \\
    CVE-2025-0851 & 022 & djl \\
    CVE-2025-53622 & 022 & DSpace \\
    CVE-2025-24790 & 276 & snowflake-jdbc \\
    CVE-2025-1391 & 284 & keycloak \\
    CVE-2025-0604 & 287 & keycloak \\
    CVE-2025-3910 & 287 & keycloak \\
    CVE-2025-27508 & 327 & emissary \\
    CVE-2025-7365 & 346 & keycloak \\
    CVE-2025-27522 & 502 & inlong \\
    CVE-2025-27496 & 532 & snowflake-jdbc \\
    CVE-2025-4641 & 611 & webdrivermanager \\
    CVE-2025-52888 & 611 & allure2 \\
    CVE-2025-48976 & 770 & commons-fileupload \\
    CVE-2025-26074 & 078 & conductor \\
    CVE-2025-23026 & 079 & jte \\
    CVE-2025-27497 & 835 & OpenDJ \\
    CVE-2025-27427 & 863 & activemq-artemis \\
    CVE-2025-30067 & 094 & kylin \\
\end{longtable}%

\section{\changed{Design Decisions}}\label{sec:appendix-design}
\vspace{-.2in}
\changed{\subsection{Web Search Tool Usage}}
\changed{In section 3.4 we discuss how free web search for vulnerability patterns was a tool we decided not to use. Free web search had less constraints in the context window. Web search lookups could fail due to web requests potentially dropping and we did not want to risk using up the context window from reliability issues. Web search also led to the agent searching for things that potentially weren't related to the CVE. We only want the agent to analyze the CVE description to better inform the vulnerability pattern. Although web search could open up the agent to more resources, we also were concerned about using too many tokens in the context window thus preferred cacheing CVE entries from the NVD database instead of performing a web search for every analysis call for the same CVE. Below we have provided transcripts for the vulnerability analysis phase for the same CVE (CVE-2025-27136) - the first example is where Claude Code's Web fetch tool \citep{claude-web-fetch} and Web search tool \citep{claude-web-search} are enabled, thus the agent can use these tools for free web search. The second example is where Web fetch and Web search are not enabled. The second example's configuration of having no free web search is what we use in \tool - we cache the CVE entries from the NVD website and we cache CWE data from MITRE in the vector database.}

\begin{lstlisting}[
    basicstyle={\scriptsize\tt},
    language={},
    frame=single,
    caption={\changed{Web search and Web fetch tools are enabled for Claude Code to use during the vulnerability analysis phase. Note: we truncated the analysis and are only showing the tool usage, tokens used, and the web links searched.}},
    label={lst:websearchtrajectory}]
Turn 5 - Assistant
Tool: WebSearch
Query: CVE-2025-27136 vulnerability details XML external entity XXE
Tokens: Input: 5 | Cache Read: 29,596 | Cache Create: 267 | Output: 67
Result: Web search results for query: "CVE-2025-27136 vulnerability 
details XML external entity XXE"

I'll search for information about the CVE-2025-27136 vulnerability and its 
relation to XML external entity (XXE) issues.

Links: [{"title":"LocalS3 CreateBucketConfiguration Endpoint XML External Entity 
(XXE) Injection . CVE-2025-27136 . GitHub Advisory Database . GitHub",
"url":"https://github.com/advisories/GHSA-g6wm-2v64-wq36"},
{"title":"CVE-2025-27136 Impact, Exploitability, and Mitigation Steps | Wiz",
"url":"https://www.wiz.io/vulnerability-database/cve/cve-2025-27136"},
{"title":"NVD - CVE-2025-27136",
"url":"https://nvd.nist.gov/vuln/detail/CVE-2025-27136"},
{"title":"XML External Entity (XXE) Processing | OWASP Foundation",
"url":"https://owasp.org/www-community/vulnerabilities/XML_External_Entity_(XXE)_
Processing"},{"title":"What is XXE (XML External Entity) | Examples & Prevention |
Imperva","url":"https://www.imperva.com/learn/application-security/
xxe-xml-external-entity/"},{"title":"CVE - Search Results",
"url":"https://cve.mitre.org/cgi-bin/cvekey.cgi?keyword=xxe"},
{"title":"What is XXE (XML external entity) injection? Tutorial & Examples
| Web Security Academy","url":"https://portswigger.net/web-security/xxe"},
{"title":"XXE Complete Guide: Impact, Examples, and Prevention | HackerOne",
"url":"https://www.hackerone.com/knowledge-center/
xxe-complete-guide-impact-examples-and-prevention"},{"title":"XML External 
Entity (XXE)","url":"https://www.invicti.com/learn/xml-external-entity-xxe"},
{"title":"XML External Entity Prevention - OWASP Cheat Sheet Series","url":
"https://cheatsheetseries.owasp.org/cheatsheets/
XML_External_Entity_Prevention_Cheat_Sheet.html"}]

Turn 6 - Assistant 
Tool: WebSearch
Query: XML external entity XXE CodeQL detection patterns Java XMLInputFactory
Tokens: Input: 4 | Cache Read: 30,998 | Cache Create: 267 | Output: 64
Result: Web search results for query: "XML external entity XXE CodeQL detection patterns Java XMLInputFactory"

I'll search for information about XML external entity (XXE) CodeQL detection patterns for Java XMLInputFactory.

Links: [{"title":"XML External Entity Prevention - OWASP Cheat Sheet Series",
"url":"https://cheatsheetseries.owasp.org/cheatsheets/XML_External_Entity_Prevention
_Cheat_Sheet.html"},{"title":"Resolving XML external entity in user-controlled data - CodeQL query help documentation","url":"https://codeql.github.com/codeql-query-help/java/java-xxe/"},
{"title":"Java false positive: XXE via XMLInputFactory . Issue #7199 . github/codeql","url":"https://github.com/github/codeql/issues/7199"},
{"title":"XML Entity Expansion in Java | SecureFlag Security Knowledge Base","url":"https://knowledge-base.secureflag.com/vulnerabilities/xml_injection/xml_entity_expansion_java.html"},
{"title":"Detecting, Fixing, and Defending Against XXE Attacks in Python and Java | by Rohit kumar | Medium","url":"https://rohitcoder.medium.com/comprehensive-guide-detecting-fixing-and-defending-against-xxe-attacks-in-python-and-java-e78691b4b918"},{"title":"Prevent XML External Entity Vulnerabilities for Java - xxe","url":"https://semgrep.dev/docs/cheat-sheets/java-xxe"},{"title":"how to fix 'Disable XML external entity (XXE) processing' vulnerabilities in java - Stack Overflow","url":"https://stackoverflow.com/questions/56777287/how-to-fix-disable-xml-external-entity-xxe-processing-vulnerabilities-in-jav"},{"title":"XXE vulnerability detection | Moderne Docs","url":"https://docs.moderne.io/openrewrite-advanced-program-analysis/security/xxe/"},{"title":"How to prevent XML external entity attack (XXE attack) - Mkyong.com","url":"https://mkyong.com/java/how-to-prevent-xml-external-entity-attack-xxe-attack/"},{"title":"xml - Java XXE vulnerability - Information Security Stack Exchange","url":"https://security.stackexchange.com/questions/260956/java-xxe-vulnerability"}]


Turn 7 - Assistant 
Tool: WebSearch
Query: CodeQL XXE taint tracking Java XML parsing sources sinks sanitizers
Tokens: Input: 3 | Cache Read: 31,265 | Cache Create: 1,354 | Output: 66
Result: Web search results for query: "CodeQL XXE taint tracking Java XML parsing sources sinks sanitizers"

I'll search for information about CodeQL XXE taint tracking in Java XML parsing, including sources, sinks, and sanitizers.

Links: [{"title":"Practical Introduction to CodeQL :: jorgectf - blog","url":"https://jorgectf.github.io/blog/post/practical-codeql-introduction/"},{"title":"Resolving XML external entity in user-controlled data - CodeQL query help documentation","url":"https://codeql.github.com/codeql-query-help/java/java-xxe/"},{"title":"Using flow state for precise data flow analysis - CodeQL","url":"https://codeql.github.com/docs/codeql-language-guides/using-flow-labels-for-precise-data-flow-analysis/"},
{"title":"CodeQL 2.16.4 (2024-03-11) CodeQL","url":"https://codeql.github.com/docs/codeql-overview/codeql-changelog/codeql-cli-2.16.4/"},{"title":"Analyzing data flow in Java and Kotlin  CodeQL","url":"https://codeql.github.com/docs/codeql-language-guides/analyzing-data-flow-in-java/"},{"title":"RealworldCTF 2024 Protected-by-Java-SE 
Writeup","url":"https://intrigus.org/research/2025/05/07/realworld-ctf-2024-using-codeql-to-find-bugs-in-codeql/"},{"title":"New dataflow API for writing custom CodeQL queries - GitHub 
Changelog","url":"https://github.blog/changelog/2023-08-14-new-dataflow-api-for-writing-custom-codeql-queries/"},
{"title":"TaintTracking - CodeQL - GitHub","url":"https://codeql.github.com/codeql-standard-libraries/javascript/semmle/javascript/dataflow/TaintTracking.qll/module.TaintTracking.html"},{"title":"How codeql deal with java's interface invoke in Global Taint Tracking . github/codeql . Discussion #7363","url":"https://github.com/github/codeql/discussions/7363"},{"title":"CodeQL 2.11.0 (2022-09-28) - CodeQL","url":"https://codeql.github.com/docs/codeql-overview/codeql-changelog/codeql-cli-2.11.0/"}]}
\end{lstlisting}

\begin{lstlisting}[
    language={},
    frame=single,
    caption={\changed{
        Web search and Web fetch tools are disabled for Claude Code to use during the vulnerability analysis phase. Claude Code can use cached data related to vulnerability patterns stored in the vector database. We have only provided the tool calls and token usage.
    }},
    label={lst:websearchexample-2}
]
Turn 4 - Assistant
Tool:chroma_get_documents(collection_name:nist_cve_2025_brief,where: 
{"cve_id": "CVE-2025-27136"})
Tokens: Input: 6 | Cache Read: 28,074 | Cache Create: 340 | Output: 105

Turn 6 - Assistant 
Tool: chroma_query_documents(collection_name: cwe_data, query_texts: 
['CWE-611', 'XML External Entity', 'XXE injection'],n_results: 3)
Tokens: Input: 5 | Cache Read: 29,031 | Cache Create: 340 | Output: 121
\end{lstlisting}
\section{CodeQL Language Server via MCP}
\label{sec:appendix-codeql-lsp}

The following are the MCP tool specifications and example usage for our custom CodeQL LSP client, wrapped as an MCP server. The MCP server code can be found at \url{https://github.com/neuralprogram/codeql-lsp-mcp}.

\subsection{Tool Specifications}\label{sec:appendix-codeql-mcp-tool-specs}

\paragraph{\texttt{codeql\_complete}}
Provides code completions at a specific position in a CodeQL file. Supports pagination for large completion lists and trigger character-based completions.

\textbf{Inputs:}
\begin{itemize}
\item \texttt{file\_uri} (string): The URI of the CodeQL file
\item \texttt{line} (number): Line number (0-based)
\item \texttt{character} (number): Character position in the line (0-based)
\item \texttt{trigger\_character} (string, optional): Optional trigger character (e.g., ".", "::")
\item \texttt{limit} (number, optional): Maximum number of completion items to return (default: 50)
\item \texttt{offset} (number, optional): Starting position for pagination (default: 0)
\end{itemize}

\textbf{Returns:} \texttt{CompletionList} with pagination metadata containing completion items, each with label, kind, documentation, and text edit information.

\noindent Example usage:
\begin{verbatim}
{
  "tool": "codeql_complete",
  "arguments": {
    "file_uri": "file:///workspace/security-query.ql",
    "line": 5,
    "character": 12,
    "trigger_character": ".",
    "limit": 25
  }
}
\end{verbatim}

\paragraph{\texttt{codeql\_hover}}
Retrieves hover information (documentation, type information) at a specific position. Provides rich markdown documentation for CodeQL predicates, classes, and modules.

\textbf{Inputs:}
\begin{itemize}
\item \texttt{file\_uri} (string): The URI of the CodeQL file
\item \texttt{line} (number): Line number (0-based)
\item \texttt{character} (number): Character position in the line (0-based)
\end{itemize}

\textbf{Returns:} \texttt{Hover | null} containing documentation content in markdown or plain text format, with optional range highlighting.

\noindent Example usage:
\begin{verbatim}
{
  "tool": "codeql_hover",
  "arguments": {
    "file_uri": "file:///workspace/security-query.ql",
    "line": 8,
    "character": 15
  }
}
\end{verbatim}

\paragraph{\texttt{codeql\_definition}}
Navigates to the definition location for a symbol at a specific position. Supports both single definitions and multiple definition locations.

\textbf{Inputs:}
\begin{itemize}
\item \texttt{file\_uri} (string): The URI of the CodeQL file
\item \texttt{line} (number): Line number (0-based)
\item \texttt{character} (number): Character position in the line (0-based)
\end{itemize}

\textbf{Returns:} \texttt{Location | Location[] | null} containing URI and range information for definition locations.

\noindent Example usage:
\begin{verbatim}
{
  "tool": "codeql_definition",
  "arguments": {
    "file_uri": "file:///workspace/security-query.ql",
    "line": 12,
    "character": 8
  }
}
\end{verbatim}

\paragraph{\texttt{codeql\_references}}
Finds all references to a symbol at a specific position across the workspace. Includes both usage references and declaration references.

\textbf{Inputs:}
\begin{itemize}
\item \texttt{file\_uri} (string): The URI of the CodeQL file
\item \texttt{line} (number): Line number (0-based)
\item \texttt{character} (number): Character position in the line (0-based)
\end{itemize}

\textbf{Returns:} \texttt{Location[] | null} containing an array of all reference locations with URI and range information.

\noindent Example usage:
\begin{verbatim}
{
  "tool": "codeql_references",
  "arguments": {
    "file_uri": "file:///workspace/security-query.ql",
    "line": 6,
    "character": 20
  }
}
\end{verbatim}

\paragraph{\texttt{codeql\_diagnostics}}
Retrieves diagnostics (errors, warnings, information messages) for a CodeQL file. Provides real-time syntax and semantic analysis results.

\textbf{Inputs:}
\begin{itemize}
\item \texttt{file\_uri} (string): The URI of the CodeQL file
\end{itemize}

\textbf{Returns:} \texttt{Diagnostic[]} containing an array of diagnostic objects with severity, message, range, and optional related information.

\noindent Example usage:
\begin{verbatim}
{
  "tool": "codeql_diagnostics",
  "arguments": {
    "file_uri": "file:///workspace/security-query.ql"
  }
}
\end{verbatim}

\paragraph{\texttt{codeql\_format}}
Formats a CodeQL file or a specific selection within the file according to CodeQL style guidelines.

\textbf{Inputs:}
\begin{itemize}
\item \texttt{file\_uri} (string): The URI of the CodeQL file
\item \texttt{range} (Range, optional): Optional range to format with start and end positions
\end{itemize}

\textbf{Returns:} \texttt{TextEdit[]} containing an array of text edits that describe the formatting changes to be applied.

\noindent Example usage:
\begin{verbatim}
{
  "tool": "codeql_format",
  "arguments": {
    "file_uri": "file:///workspace/security-query.ql",
    "range": {
      "start": { "line": 10, "character": 0 },
      "end": { "line": 25, "character": 0 }
    }
  }
}
\end{verbatim}

\paragraph{\texttt{codeql\_update\_file}}
Updates the content of an open CodeQL file in the language server. This allows for dynamic content modification and analysis of unsaved changes.

\textbf{Inputs:}
\begin{itemize}
\item \texttt{file\_uri} (string): The URI of the CodeQL file
\item \texttt{content} (string): The new complete content of the file
\end{itemize}

\textbf{Returns:} \texttt{string} containing a success confirmation message.

\noindent Example usage:
\begin{verbatim}
{
  "tool": "codeql_update_file",
  "arguments": {
    "file_uri": "file:///workspace/security-query.ql",
    "content": "import cpp\n\nfrom Function f\nwhere f.hasName(\"strcpy\")\nselect f, \"Unsafe string copy function\""
  }
}
\end{verbatim}
\section{CodeQL Queries}
\subsection{CodeQL Query Structure Template}
\label{sec:appendix-query-structure}
The template below is given to the LLM agent at the start of the iterative query synthesis task. The prompt instructs the LLM to use the AST nodes, along with the CodeQL LSP and CodeQL references in the vector database, to fill in this template. The prompt also takes note to find similar queries related to the given CVE's vulnerability.
\vspace{.2in}
\begin{lstlisting}[language=Java,numbers=left,basicstyle=\ttfamily\footnotesize]
/**
 * @name [Vulnerability Name based on analysis]
 * @description [Description derived from the vulnerability pattern]
 * @problem.severity error
 * @security-severity [score based on severity]
 * @precision high 
 * @tags security
 * @kind path-problem
 * @id [unique-id]
 */
import java
import semmle.code.java.frameworks.Networking
import semmle.code.java.dataflow.DataFlow
import semmle.code.java.dataflow.FlowSources
import semmle.code.java.dataflow.TaintTracking
private import semmle.code.java.dataflow.ExternalFlow

class Source extends DataFlow::Node {
  Source() {
    exists([AST node type from analysis] |
      /* Fill based on AST patterns for sources identified in Phase 1 & 2 */ 
      and this.asExpr() = [appropriate mapping]
    )
  }
}
 
class Sink extends DataFlow::Node {
  Sink() {
    exists([AST node type] | 
      /* Fill based on AST patterns for sinks */  
      and this.asExpr() = [appropriate mapping]
    ) or
    exists([Alternative AST pattern] |  
      /* Additional sink patterns from analysis */ 
      and [appropriate condition]
    )
  }
}

class Sanitizer extends DataFlow::Node {
  Sanitizer() {
    exists([AST node type for sanitizers] |
      /* Fill based on sanitizer patterns from Phase 1 & 2 */ 
    )
  }
}

module MyPathConfig implements DataFlow::ConfigSig {
  predicate isSource(DataFlow::Node source) {
    source instanceof Source  
  }

  predicate isSink(DataFlow::Node sink) {
    sink instanceof Sink
  }

  predicate isBarrier(DataFlow::Node sanitizer) {
    sanitizer instanceof Sanitizer
  }

  predicate isAdditionalFlowStep(DataFlow::Node n1, DataFlow::Node n2) {{
    /* Fill based on additional taint steps from analysis */
  }}
}}

module MyPathFlow = TaintTracking::Global<MyPathConfig>;
import MyPathFlow::PathGraph

from
  MyPathFlow::PathNode source,
  MyPathFlow::PathNode sink
where
  MyPathFlow::flowPath(source, sink) 
select
  sink.getNode(),
  source,
  sink,
  "[Alert message based on vulnerability]",
  source.getNode(),
  "[source description]"
\end{lstlisting}

\subsection{Iterative Feedback Example}\label{sec:appendix-iterative-feedback}
The following is an excerpt from the prompt used in a new context window that instructs the agent to improve the last iteration's query. The feedback is derived from automatically running synthesized queries against the fixed and vulnerable versions of the CVE's repository, and comparing the query's results against the changed methods in the CVE fix commit diff. The feedback also includes any successful method hits, false positive method hits, and which fixed methods have not been detected by the query. 

\begin{lstlisting}[language=Java,numbers=left,basicstyle=\ttfamily\footnotesize]
## Objective
Refine the CodeQL query based on previous iteration feedback to improve vulnerability detection.

## Previous Iteration Feedback
## Iteration 2 Results

## Previous Query (Iteration 2)
```ql
/**
 * @name AntiSamy XSS Bypass in Style Tags
 * @description Detects potential XSS vulnerabilities where AntiSamy HTML sanitization fails to properly filter malicious style tags with event handlers
 * @problem.severity error
 * @security-severity 6.1
 * @precision high
 * @tags security
 * @kind path-problem
 * @id java/antisamy-style-xss-bypass
 */
import java
import semmle.code.java.dataflow.DataFlow
import semmle.code.java.dataflow.TaintTracking

class MaliciousHtmlInput extends DataFlow::Node {
  MaliciousHtmlInput() {
    exists(StringLiteral sl |
      this.asExpr() = sl and
      (
        sl.getValue().matches("%<style%onload%>%") or
        sl.getValue().matches("%<style%onclick%>%") or
        sl.getValue().matches("%<style%onerror%>%") or
        sl.getValue().matches("%<style%onmouseover%>%") or
        sl.getValue().matches("%style%alert%") or
        sl.getValue().matches("%<style%javascript%") or
        sl.getValue().matches("%<style%expression%")
      )
    ) or
    exists(Parameter p |
      this.asParameter() = p and
      (
        p.getName().toLowerCase().matches("%html%") or
        p.getName().toLowerCase().matches("%input%") or
        p.getName().toLowerCase().matches("%content%")
      )
    )
  }
}

class AntiSamyProcessing extends DataFlow::Node {
  AntiSamyProcessing() {
    exists(MethodCall mc |
      this.asExpr() = mc and
      mc.getMethod().hasName("scan") and
      mc.getMethod().getDeclaringType()(RefType)
      .hasQualifiedName("org.owasp.validator.html", "AntiSamy")
    ) or
    exists(MethodCall mc |
      this.asExpr() = mc and
      mc.getMethod().hasName("getCleanHTML") and
      mc.getQualifier().getType().(RefType)
      .hasQualifiedName("org.owasp.validator.html", "CleanResults")
    )
  }
}

class PotentialXssOutput extends DataFlow::Node {
  PotentialXssOutput() {
    exists(MethodCall mc |
      this.asExpr() = mc and
      (
        mc.getMethod().hasName("contains") and
        exists(StringLiteral sl |
          mc.getAnArgument() = sl and
          (
            sl.getValue().matches("%alert%") or
            sl.getValue().matches("%script%") or
            sl.getValue().matches("%javascript%") or
            sl.getValue().matches("%onload%") or
            sl.getValue().matches("%onclick%")
          )
        )
      )
    ) or
    exists(MethodCall mc |
      this.asExpr() = mc and
      mc.getMethod().hasName("getCleanHTML") and
      mc.getQualifier().getType().(RefType)
      .hasQualifiedName("org.owasp.validator.html", "CleanResults")
    )
  }
}

class ImproperSanitizationBarrier extends DataFlow::Node {
  ImproperSanitizationBarrier() {
    exists(MethodCall mc |
      this.asExpr() = mc and
      mc.getMethod().hasName("assertTrue") and
      exists(LogNotExpr lne |
        mc.getAnArgument() = lne and
        lne.getExpr().(MethodCall).getMethod().hasName("contains")
      )
    )
  }
}

module AntiSamyXssConfig implements DataFlow::ConfigSig {
  predicate isSource(DataFlow::Node source) {
    source instanceof MaliciousHtmlInput
  }

  predicate isSink(DataFlow::Node sink) {
    sink instanceof PotentialXssOutput
  }

  predicate isBarrier(DataFlow::Node sanitizer) {
    sanitizer instanceof ImproperSanitizationBarrier
  }

  predicate isAdditionalFlowStep(DataFlow::Node n1, DataFlow::Node n2) {
    exists(MethodCall mc |
      n1.asExpr() = mc.getQualifier() and
      n2.asExpr() = mc and
      mc.getMethod().hasName("scan") and
      mc.getMethod().getDeclaringType().(RefType)
      .hasQualifiedName("org.owasp.validator.html", "AntiSamy")
    ) or
    exists(MethodCall mc |
      n1.asExpr() = mc.getQualifier() and
      n2.asExpr() = mc and
      mc.getMethod().hasName("getCleanHTML")
    )
  }
}

module AntiSamyXssFlow = TaintTracking::Global<AntiSamyXssConfig>;
import AntiSamyXssFlow::PathGraph

from
  AntiSamyXssFlow::PathNode source,
  AntiSamyXssFlow::PathNode sink
where
  AntiSamyXssFlow::flowPath(source, sink)
select
  sink.getNode(),
  source,
  sink,
  "Potential XSS vulnerability: HTML input with malicious style tags may bypass AntiSamy sanitization",
  source.getNode(),
  "malicious HTML input"
```

## Compilation Results
COMPILATION SUCCESS: Query syntax is valid

## Execution Results
## Query Evaluation Summary (Iteration 2)
Results: Vulnerable=8, Fixed=8
Method Recall: Vulnerable=True, Fixed=True
True Positive Methods: Vulnerable=2, Fixed=2
Coverage: 1/1 target methods
PARTIAL: Query hits targets but has false positives in fixed version
Method location format is path/to/hit/file.java:[Class of hit method]:[Hit method]
Successfully targeted methods:
  -src/main/java/org/owasp/validator/html/scan/MagicSAXFilter.java:MagicSAXFilter:startElement
False positives (hits in fixed version):
  - src/main/java/org/owasp/validator/html/scan/MagicSAXFilter.java:MagicSAXFilter:startElement

## Detailed Evaluation Analysis

**Method Coverage**: 1/1 target methods
**File Coverage**: 1/1 target files
**Successfully targeted files**:
   - MagicSAXFilter.java

**Successfully targeted methods**:
   - startElement

**False positives (hit in fixed version)**:
   - startElement

**Evaluation Summary**:
- Vulnerable DB File Recall: True
- Fixed DB File Recall: True
- Vulnerable DB Method Recall: True
- Fixed DB Method Recall: True
- Total Query Results: Vulnerable=8, Fixed=8
- Code Flow Paths: Vulnerable=8, Fixed=8

## Next Steps
We want vulnerable DB method recall and we don't want fixed DB method recall!
**Priority**: Reduce false positives by adding more specific conditions to avoid hitting the methods listed above.
\end{lstlisting}

\subsection{\tool Synthesized Query Examples}\label{sec:appendix-query-examples}
Below are examples of successful synthesized queries. The queries successfully find $>$ 0 true positive method hits on the vulnerable CodeQL database of the CVE's source code, and no false positive method hits on the fixed version's CodeQL database. For reference queries to compare with, CWE queries can be found on the official CodeQL repository \citep{noauthor_codeqljavaqlsrcsecuritycwe_nodate, noauthor_codeqljavaqllibsemmlecodejavasecurity_nodate} \\ 

\textbf{CVE-2025-27136, CWE-611 - Improper Restriction of XML External Entity Reference } \\
\vspace{-.2in}
\begin{lstlisting}[language=Java,numbers=left,basicstyle=\ttfamily\footnotesize]
/**
 * @name XML External Entity vulnerability in WstxInputFactory without secure configuration
 * @description WstxInputFactory used in XmlFactory without disabling DTD support and external entities allows XXE attacks
 * @problem.severity error
 * @security-severity 9.1
 * @precision high
 * @tags security
 * @kind path-problem
 * @id java/wstxinputfactory-xxe
 */

import java
import semmle.code.java.dataflow.DataFlow
import semmle.code.java.dataflow.FlowSources
import semmle.code.java.dataflow.TaintTracking
private import semmle.code.java.dataflow.ExternalFlow

class WstxInputFactoryCreation extends DataFlow::Node {
  WstxInputFactoryCreation() {
    exists(MethodCall mc |
      mc.getMethod().hasQualifiedName("com.ctc.wstx.stax", "WstxInputFactory", "newInstance") or
      mc.getMethod().hasQualifiedName("com.ctc.wstx.stax", "WstxInputFactory", "newFactory")
    |
      this.asExpr() = mc
    ) or
    exists(ClassInstanceExpr cie |
      cie.getConstructedType().hasQualifiedName("com.ctc.wstx.stax", "WstxInputFactory") and
      this.asExpr() = cie
    ) or
    // Include variable access to WstxInputFactory instances (like "input" parameter)
    exists(Variable v, VarAccess va |
      v.getType().(RefType).hasQualifiedName("com.ctc.wstx.stax", "WstxInputFactory") and
      va.getVariable() = v and
      this.asExpr() = va
    )
  }
}

class UnsafeXmlFactoryUsage extends DataFlow::Node {
  UnsafeXmlFactoryUsage() {
    exists(ClassInstanceExpr xmlFactoryCall |
      // XmlFactory constructor with WstxInputFactory parameter
      xmlFactoryCall.getConstructedType()
      .hasQualifiedName("com.fasterxml.jackson.dataformat.xml", "XmlFactory") and
      xmlFactoryCall.getArgument(0) = this.asExpr()
    ) or
    exists(ClassInstanceExpr xmlMapperCall, ClassInstanceExpr xmlFactoryCall |
      // XmlMapper constructor using XmlFactory with WstxInputFactory
      xmlMapperCall.getConstructedType()
      .hasQualifiedName("com.fasterxml.jackson.dataformat.xml", "XmlMapper") and
      xmlFactoryCall.getConstructedType()
      .hasQualifiedName("com.fasterxml.jackson.dataformat.xml", "XmlFactory") and
      xmlMapperCall.getArgument(0) = xmlFactoryCall and
      xmlFactoryCall.getArgument(0) = this.asExpr()
    )
  }
}

class WstxInputFactorySanitizer extends DataFlow::Node {
  WstxInputFactorySanitizer() {
    exists(MethodCall setPropertyCall, VarAccess factoryVar |
      setPropertyCall.getMethod().hasQualifiedName("javax.xml.stream", "XMLInputFactory", "setProperty") and
      setPropertyCall.getQualifier() = factoryVar and
      (
        // DTD support disabled
        (exists(Field f |
          setPropertyCall.getArgument(0) = f.getAnAccess() and
          f.hasName("SUPPORT_DTD") and
          f.getDeclaringType().hasQualifiedName("javax.xml.stream", "XMLInputFactory")
        ) and
        exists(Field f |
          setPropertyCall.getArgument(1) = f.getAnAccess() and
          f.hasName("FALSE") and
          f.getDeclaringType().hasQualifiedName("java.lang", "Boolean")
        )) or
        // External entities disabled
        (exists(Field f |
          setPropertyCall.getArgument(0) = f.getAnAccess() and
          f.hasName("IS_SUPPORTING_EXTERNAL_ENTITIES") and
          f.getDeclaringType().hasQualifiedName("javax.xml.stream", "XMLInputFactory")
        ) and
        exists(Field f |
          setPropertyCall.getArgument(1) = f.getAnAccess() and
          f.hasName("FALSE") and
          f.getDeclaringType().hasQualifiedName("java.lang", "Boolean")
        ))
      ) and
      this.asExpr() = factoryVar
    )
  }
}

module WstxInputFactoryFlowConfig implements DataFlow::ConfigSig {
  predicate isSource(DataFlow::Node source) {
    source instanceof WstxInputFactoryCreation
  }

  predicate isSink(DataFlow::Node sink) {
    sink instanceof UnsafeXmlFactoryUsage
  }

  predicate isBarrier(DataFlow::Node sanitizer) {
    sanitizer instanceof WstxInputFactorySanitizer
  }

  predicate isAdditionalFlowStep(DataFlow::Node n1, DataFlow::Node n2) {
    // WstxInputFactory passed to XmlFactory constructor
    exists(ClassInstanceExpr xmlFactoryCall |
      xmlFactoryCall.getConstructedType()
      .hasQualifiedName("com.fasterxml.jackson.dataformat.xml", "XmlFactory") and
      xmlFactoryCall.getArgument(0) = n1.asExpr() and
      n2.asExpr() = xmlFactoryCall
    ) or
    // XmlFactory passed to XmlMapper constructor
    exists(ClassInstanceExpr xmlMapperCall |
      xmlMapperCall.getConstructedType()
      .hasQualifiedName("com.fasterxml.jackson.dataformat.xml", "XmlMapper") and
      xmlMapperCall.getArgument(0) = n1.asExpr() and
      n2.asExpr() = xmlMapperCall
    )
  }
}

module WstxInputFactoryFlow = TaintTracking::Global<WstxInputFactoryFlowConfig>;
import WstxInputFactoryFlow::PathGraph

from
  WstxInputFactoryFlow::PathNode source,
  WstxInputFactoryFlow::PathNode sink
where
  WstxInputFactoryFlow::flowPath(source, sink)
select
  sink.getNode(),
  source,
  sink,
  "WstxInputFactory used without secure configuration flows to XML parser, allowing XXE attacks",
  source.getNode(),
  "WstxInputFactory usage"
\end{lstlisting}
\textbf{CVE-2025-0851, CWE-22 - Path Traversal}

\begin{lstlisting}[language=Java,numbers=left,basicstyle=\ttfamily\footnotesize]
/**
 * @name Archive path traversal vulnerability (ZipSlip) - CVE-2025-0851
 * @description Archive entries with path traversal sequences can write files outside the intended extraction directory
 * @problem.severity error
 * @security-severity 9.8
 * @precision high
 * @tags security
 * @kind path-problem
 * @id java/archive-path-traversal-cve-2025-0851
 */

import java
import semmle.code.java.dataflow.DataFlow
import semmle.code.java.dataflow.TaintTracking

/**
 * Sources: Archive entry names from ZipEntry.getName() and TarArchiveEntry.getName()
 */
class ArchiveEntryNameSource extends DataFlow::Node {
  ArchiveEntryNameSource() {
    exists(MethodCall mc |
      mc.getMethod().getName() = "getName" and    (mc.getMethod().getDeclaringType().hasQualifiedName("java.util.zip", "ZipEntry") or mc.getMethod()
            .getDeclaringType()          .hasQualifiedName("org.apache.commons.compress.archivers.tar", "TarArchiveEntry")
      ) and
      this.asExpr() = mc
    )
  }
}

/**
 * Sinks: Path resolution operations that lead to file creation
 */
class PathCreationSink extends DataFlow::Node {
  PathCreationSink() {
    // Arguments to Path.resolve() calls
    exists(MethodCall resolveCall |
      resolveCall.getMethod().getName() = "resolve" and
      resolveCall.getMethod().getDeclaringType()
      .hasQualifiedName("java.nio.file", "Path") and
      this.asExpr() = resolveCall.getAnArgument()
    )
    or
    // Arguments to file creation operations
    exists(MethodCall fileOp |
      (
        fileOp.getMethod().getName() = "createDirectories" or
        fileOp.getMethod().getName() = "newOutputStream" or
        fileOp.getMethod().getName() = "write" or
        fileOp.getMethod().getName() = "copy"
      ) and
      fileOp.getMethod().getDeclaringType().hasQualifiedName("java.nio.file", "Files") and
      this.asExpr() = fileOp.getAnArgument()
    )
  }
}

/**
 * Sanitizers: Proper validation that prevents path traversal
 */
class PathTraversalSanitizer extends DataFlow::Node {
  PathTraversalSanitizer() {
    // The validateArchiveEntry method call that properly validates paths
    // This blocks flow after the validation call is made
    exists(MethodCall validateCall |
      validateCall.getMethod().getName() = "validateArchiveEntry" and
      (
        // Any variable assigned from validateArchiveEntry call result
        exists(Variable v |
          this.asExpr() = v.getAnAccess() and
          exists(AssignExpr assign |
            assign.getDest() = v.getAnAccess() and
            assign.getRhs() = validateCall
          )
        )
        or
        // Variables passed through validateArchiveEntry calls
        this.asExpr() = validateCall.getAnArgument() and
        exists(ExprStmt stmt | stmt.getExpr() = validateCall)
      )
    )
    or
    // Proper ".." validation with exception throwing (complete pattern)
    exists(MethodCall containsCall, IfStmt ifStmt, ThrowStmt throwStmt |
      containsCall.getMethod().getName() = "contains" and
      containsCall.getAnArgument().(StringLiteral).getValue() = ".." and
      ifStmt.getCondition().getAChildExpr*() = containsCall and
      ifStmt.getThen().getAChild*() = throwStmt and
      this.asExpr() = containsCall.getQualifier()
    )
    or
    // Path normalization combined with startsWith validation
    exists(MethodCall normalizeCall, MethodCall startsWithCall |
      normalizeCall.getMethod().getName() = "normalize" and
      normalizeCall.getMethod().getDeclaringType()
      .hasQualifiedName("java.nio.file", "Path") and
      startsWithCall.getMethod().getName() = "startsWith" and
      startsWithCall.getMethod().getDeclaringType().
      hasQualifiedName("java.nio.file", "Path") and
      DataFlow::localFlow(DataFlow::exprNode(normalizeCall),
        DataFlow::exprNode(startsWithCall.getQualifier())) and
      this.asExpr() = normalizeCall.getQualifier()
    )
  }
}

/**
 * Additional predicate to detect validation barriers at method level
 */
predicate hasValidationCall(Callable method) {
  exists(MethodCall validateCall |
    validateCall.getEnclosingCallable() = method and
    validateCall.getMethod().getName() = "validateArchiveEntry"
  )
}

module PathTraversalConfig implements DataFlow::ConfigSig {
  predicate isSource(DataFlow::Node source) { 
    source instanceof ArchiveEntryNameSource 
  }

  predicate isSink(DataFlow::Node sink) { 
    sink instanceof PathCreationSink 
  }

  predicate isBarrier(DataFlow::Node sanitizer) {
    sanitizer instanceof PathTraversalSanitizer
  }

  predicate isBarrierIn(DataFlow::Node node) {
    // Barrier at method entry if method contains validateArchiveEntry call
    node instanceof DataFlow::ParameterNode and
    hasValidationCall(node.getEnclosingCallable())
  }

  predicate isAdditionalFlowStep(DataFlow::Node n1, DataFlow::Node n2) {
    // Flow through variable assignments and declarations
    exists(LocalVariableDeclExpr decl |
      decl.getInit() = n1.asExpr() and
      n2.asExpr() = decl.getVariable().getAnAccess()
    )
    or
    exists(AssignExpr assign |
      assign.getRhs() = n1.asExpr() and
      n2.asExpr() = assign.getDest()
    )
    or
    // Flow through string manipulation methods that preserve taint
    exists(MethodCall mc |
      mc.getAnArgument() = n1.asExpr() and
      n2.asExpr() = mc and
      (
        mc.getMethod().getName() = "removeLeadingFileSeparator" or
        mc.getMethod().getName() = "trim" or
        mc.getMethod().getName() = "toString" or
        mc.getMethod().getName() = "substring"
      )
    )
    or
    // Flow through Path operations
    exists(MethodCall pathOp |
      pathOp.getAnArgument() = n1.asExpr() and
      n2.asExpr() = pathOp and
      pathOp.getMethod().getName() = "resolve" and
      pathOp.getMethod().getDeclaringType()
      .hasQualifiedName("java.nio.file", "Path")
    )
  }
}

module PathTraversalFlow = TaintTracking::Global<PathTraversalConfig>;

import PathTraversalFlow::PathGraph

from PathTraversalFlow::PathNode source, PathTraversalFlow::PathNode sink
where
  PathTraversalFlow::flowPath(source, sink) and
  // Focus on the specific vulnerable files and methods
  (
    source.getNode().getEnclosingCallable().getDeclaringType()
    .hasName("TarUtils") or
    source.getNode().getEnclosingCallable().getDeclaringType()
    .hasName("ZipUtils")
  ) and
  (
    source.getNode().getEnclosingCallable().getName() = "untar" or
    source.getNode().getEnclosingCallable().getName() = "unzip"
  ) and
  // Only report flows where validation is NOT properly done
  not hasValidationCall(source.getNode().getEnclosingCallable())
select sink.getNode(), source, sink,
  "Archive entry name from $@ flows to file system operation without proper path traversal validation, allowing ZipSlip attack.",
  source.getNode(), "archive entry name"
\end{lstlisting}

\textbf{CVE-2025-27528, CWE-502 - Deserialization of Untrusted Data}
\begin{lstlisting}[language=Java,numbers=left,basicstyle=\ttfamily\footnotesize]
/**
 * @name MySQL JDBC URL parameter injection vulnerability
 * @description Detects MySQL JDBC URLs with dangerous bracket parameters that bypass inadequate filtering in vulnerable code
 * @problem.severity error
 * @security-severity 8.8
 * @precision high
 * @tags security
 * @kind path-problem
 * @id java/mysql-jdbc-url-injection
 */

import java
import semmle.code.java.dataflow.DataFlow
import semmle.code.java.dataflow.TaintTracking

class MySQLDangerousBracketUrlSource extends DataFlow::Node {
  MySQLDangerousBracketUrlSource() {
    // String literals with dangerous MySQL parameters in bracket notation
    exists(StringLiteral lit |
      lit.getValue().matches("*mysql*") and
      lit.getValue().matches("*[*]*") and
      (
        lit.getValue().matches("*allowLoadLocalInfile*") or
        lit.getValue().matches("*allowUrlInLocalInfile*") or
        lit.getValue().matches("*autoDeserialize*") or
        lit.getValue().matches("*allowPublicKeyRetrieval*") or
        lit.getValue().matches("*serverTimezone*") or
        lit.getValue().matches("*user*") or
        lit.getValue().matches("*password*")
      ) and
      this.asExpr() = lit
    )
    or
    // Parameters to filterSensitive method that may contain dangerous bracket content
    exists(Method m, Parameter p |
      m.hasName("filterSensitive") and
      m.getDeclaringType().getName() = "MySQLSensitiveUrlUtils" and
      p = m.getAParameter() and
      this.asParameter() = p
    )
  }
}

class VulnerableCodePatternSink extends DataFlow::Node {
  VulnerableCodePatternSink() {
    // The vulnerability: calls to filterSensitive in vulnerable code patterns
    exists(Method m, MethodCall filterCall |
      m.hasName("filterSensitive") and
      m.getDeclaringType().getName() = "MySQLSensitiveUrlUtils" and
      filterCall.getMethod() = m and
      this.asExpr() = filterCall and
      // Key vulnerability condition: this code exists where filterSensitiveKeyByBracket method is NOT available
      // In the vulnerable version, filterSensitiveKeyByBracket doesn't exist
      not exists(Method bracketMethod |
        bracketMethod.hasName("filterSensitiveKeyByBracket") and
        bracketMethod.getDeclaringType().getName() = "MySQLSensitiveUrlUtils" and
        bracketMethod.getDeclaringType() = m.getDeclaringType()
      )
    )
    or
    // Additional sink: method calls that use the result of inadequate filtering
    exists(MethodCall mc, MethodCall filterCall |
      filterCall.getMethod().hasName("filterSensitive") and
      filterCall.getMethod().getDeclaringType().getName() = "MySQLSensitiveUrlUtils" and
      DataFlow::localFlow(DataFlow::exprNode(filterCall), DataFlow::exprNode(mc.getArgument(_))) and
      this.asExpr() = mc and
      // Only vulnerable if no proper bracket filtering exists in the same class
      not exists(Method bracketMethod |
        bracketMethod.hasName("filterSensitiveKeyByBracket") and
        bracketMethod.getDeclaringType().getName() = "MySQLSensitiveUrlUtils" and
        bracketMethod.getDeclaringType() = filterCall.getMethod().getDeclaringType()
      )
    )
  }
}

class ProperBracketFilteringSanitizer extends DataFlow::Node {
  ProperBracketFilteringSanitizer() {
    // The proper bracket-based sanitization method (present only in fixed version)
    exists(MethodCall mc |
      mc.getMethod().hasName("filterSensitiveKeyByBracket") and
      mc.getMethod().getDeclaringType().getName() = "MySQLSensitiveUrlUtils" and
      this.asExpr() = mc
    )
  }
}

module MySQLJDBCUrlInjectionConfig implements DataFlow::ConfigSig {
  predicate isSource(DataFlow::Node source) {
    source instanceof MySQLDangerousBracketUrlSource
  }

  predicate isSink(DataFlow::Node sink) {
    sink instanceof VulnerableCodePatternSink
  }

  predicate isBarrier(DataFlow::Node sanitizer) {
    sanitizer instanceof ProperBracketFilteringSanitizer
  }

  predicate isAdditionalFlowStep(DataFlow::Node n1, DataFlow::Node n2) {
    // Flow through string concatenation operations
    exists(AddExpr addExpr |
      n1.asExpr() = addExpr.getLeftOperand() and
      n2.asExpr() = addExpr
    )
    or
    exists(AddExpr addExpr |
      n1.asExpr() = addExpr.getRightOperand() and
      n2.asExpr() = addExpr
    )
    or
    // Flow through variable assignments
    exists(Assignment assign |
      n1.asExpr() = assign.getSource() and
      n2.asExpr() = assign.getDest()
    )
    or
    // Flow through return statements
    exists(ReturnStmt ret |
      n1.asExpr() = ret.getResult() and
      n2.asParameter() = ret.getEnclosingCallable().getAParameter()
    )
  }
}

module MySQLJDBCUrlInjectionFlow = TaintTracking::Global<MySQLJDBCUrlInjectionConfig>;

import MySQLJDBCUrlInjectionFlow::PathGraph

from MySQLJDBCUrlInjectionFlow::PathNode source, MySQLJDBCUrlInjectionFlow::PathNode sink
where MySQLJDBCUrlInjectionFlow::flowPath(source, sink)
select sink.getNode(), source, sink,
  "MySQL JDBC URL with dangerous bracket parameters flows to vulnerable filtering logic at $@ that lacks proper bracket-based sanitization",
  source.getNode(), "dangerous URL source"
\end{lstlisting}

\subsection{AST Extraction Query}\label{sec:appendix-ast-query}
Given a fix diff, \tool automatically parses the changed methods and files, and inserts them into an AST pretty printing query template. Below is an example of the AST extraction query used for CVE-2014-7816.

\begin{lstlisting}[language=Java,numbers=left,basicstyle=\ttfamily\footnotesize]
/**
 * @name Expressions and statements for CVE-2014-7816 changed code areas
 * @description Extract expressions and statements from vulnerability fix areas
 * @id java/expr-stmt-diff-CVE_2014_7816
 * @kind problem
 * @problem.severity recommendation
 */

import java

from Element e, Location l
where 
  l = e.getLocation() and ((l.getFile().getBaseName() = "PathSeparatorHandler.java" 
  and l.getStartLine() >= 1 and l.getEndLine() <= 100) or 
  
  (l.getFile().getBaseName() = "URLDecodingHandler.java" 
  and l.getStartLine() >= 17 and l.getEndLine() <= 128)
  
  or (l.getFile().getBaseName() = "ResourceHandler.java" 
  and l.getStartLine() >= 158 and l.getEndLine() <= 172) 
  
  or (l.getFile().getBaseName() = "io.undertow.server.handlers.builder.HandlerBuilder" and l.getStartLine() >= 17
  and l.getEndLine() <= 29) 
  
  or (l.getFile().getBaseName() = "DefaultServlet.java" 
  and l.getStartLine() >= 39 
  and l.getEndLine() <= 150) 
  
  or (l.getFile().getBaseName() = "ServletPathMatches.java" 
  and l.getStartLine() >= 32
  and l.getEndLine() <= 140)) 
select e, 
  e.toString() as element,
  e.getAPrimaryQlClass() as elementType,
  l.getFile().getBaseName() as file,
  l.getStartLine() as startLine,
  l.getEndLine() as endLine,
  l.getStartColumn() as startColumn,
  l.getEndColumn() as endColumn
\end{lstlisting}
\end{document}